\documentclass[a4paper,11pt]{article}
\pdfoutput=1 

\usepackage{jheppub} 

\usepackage[T1]{fontenc} 
\usepackage[T1]{fontenc}
\usepackage[utf8]{inputenc}
\usepackage{epsfig}
\usepackage{graphicx}
\usepackage{latexsym}
\usepackage{textcomp}
\usepackage{amssymb}
\usepackage{amsfonts,amsthm,amstext,amscd}
\usepackage[normalem]{ulem}
\usepackage[dvipsnames]{xcolor}

\def\be{\begin{equation}}
\def\ee{\end{equation}}
\def\bea{\begin{eqnarray}}
\def\eea{\end{eqnarray}}

\def\le{\left}
\def\ri{\right}
\def\Sq{S}
\def\Sqt{\widetilde S}
\def\re{\mathrm{Re}}

\def\lab{\label}
\def\tr{\textrm{Tr}}
\title{\boldmath Analytic long-lived modes in charged critical plasma}

\author[a]{Umut G\"ursoy,}
\author[b,c]{Matti J\"arvinen,}
\author[d]{Giuseppe Policastro}
\author[a]{and Natale Zinnato}

\affiliation[a]{Institute for Theoretical Physics, Utrecht University, Leuvenlaan 4, 3584 CE Utrecht, The Netherlands}
\affiliation[b]{Asia Pacific Center for Theoretical Physics, Pohang 37673, Republic of Korea}
\affiliation[c]{Department of Physics, Pohang University of Science and Technology, Pohang 37673, Republic of Korea}
\affiliation[d]{Laboratoire de Physique de l’Ecole normale supérieure, ENS, Universit\'e PSL, CNRS, Sorbonne
Universit\'e, Universit\'e de Paris, F-75005 Paris, France}

\emailAdd{u.gursoy@uu.nl}
\emailAdd{matti.jarvinen@apctp.org}
\emailAdd{giuseppe.policastro@phys.ens.fr}
\emailAdd{n.zinnato@uu.nl}

\preprint{APCTP Pre2021 - 031}

\abstract{Fluctuations around critical behavior of a holographic charged plasmas are investigated by studying quasi-normal modes of the corresponding black branes in 5D Einstein-Maxwell-Dilaton gravity. The near horizon geometry of black branes approaches the well-known 2D charged string black hole in the critical limit, for which a world-sheet description is available, and the corresponding quasi-normal modes can be obtained analytically from the reflection amplitude of the 2D black hole geometry. We find two distinct set of modes: a purely imaginary ``decoupled'' set, directly following from the reflection amplitude, and a ``non-decoupled'' set that was already identified in the neutral holographic plasma in \cite{Betzios:2018kwn}. In the extremal limit, the former set of imaginary quasi-normal modes coalesce on a branch cut starting from the  the origin, signaling breakdown of hydrodynamic approximation. We further complete the black brane geometry with a slice of AdS near the boundary, to allow for a holographic construction, and find another set of modes localized in the UV. Finally, we develop an alternative WKB method to obtain the quasi-normal modes in the critical limit and apply this method to study the spectrum of hyperscaling-violating Lifshitz black branes.  The critical limit of the plasma we consider in this paper is in one-to-one correspondence with the large D limit of Einstein's gravity which allows for an alternative interesting interpretation of our findings. 
}

\begin{document} 
\maketitle
\flushbottom

 
\section{Introduction} 

Critical behaviour is a central theme in theoretical and experimental studies of quantum many body systems, ranging from condensed matter to high energy physics. In particular, special scaling properties of observables such as conformal or Lifshitz invariance that emerge near criticality provide strong constraints, and, in some cases allow for a full solution, as for example, in the case of the 3D Ising model \cite{El-Showk:2014dwa}. Quantum criticality at vanishing temperature has also been argued \cite{sachdev_2011} to be the key ingredient in yet unsolved problems such as the high $T_c$ superconductors. On the high energy side, critical behavior at finite temperature is important for understanding the QCD phase diagram which is believed to possess a critical point at finite temperature and baryon chemical potential \cite{Rajagopal:1992qz,Alford:1997zt,Stephanov:1998dy}; see  \cite{Rischke:2003mt} for a review.  

Critical behavior has also been the main focus of attention in applications of the AdS/CFT correspondence to strongly interacting many body systems. In particular the ${\cal N}=4$ Super Yang-Mills theory, the canonical example of conformal invariant 4D gauge theories, has been used in the large $N$ limit  as a proxy for the approximate conformal behavior observed in QCD above the confinement-deconfinement crossover and the thermodynamic, transport and dynamical properties of the quark-gluon plasma at strong coupling have been studied using 5D AdS black holes.  Similarly, construction of holographic duals of Lifshitz invariant, and hyperscaling violating theories recently played an important role in studying transport and diffusion in quantum field theories with such generalized scaling behavior; see for example \cite{Gouteraux:2011ce}. 

Unlike these standard examples where holographic descriptions exhibit scaling all the way from the UV to the IR, in a generic quantum field theory, critical scaling properties arise only in certain regions in the phase diagram whereas the theory in the vacuum state could, for example, have a gapped spectrum hence no critical behavior, QCD being an example relevant to this paper. 
Realization of this behavior, in particular presence of confinement in the vacuum and a deconfined plasma state at finite temperature, in the so-called bottom-up holographic theories using dilaton-gravity in 5D has a long history starting from \cite{Gursoy:2007cb,Gursoy:2007er,Gubser:2008ny}, see \cite{Gursoy:2021efc} and \cite{Jarvinen:2021jbd} for recent reviews\footnote{It has proven difficult to find a realization of this behavior in a consistent top-down holographic model, {\emph{e.g.}}, the GPPZ model \cite{Girardello:1999bd} that describes ${\mathcal N}=1$ SYM has a singular ten-dimensional geometry, and the Witten model \cite{Witten:1998zw} describes a higher-dimensional theory in the UV. }. See also \cite{DeWolfe:2010he} and \cite{Alho:2013hsa} for inclusion of baryon charge in these holographic models. The former also analyzed the thermodynamics near the QCD critical point in this holographic setting.

However, the emergence of critical behavior in these ``confining'' holographic theories has only recently been addressed \cite{Gursoy:2015nza,Betzios:2017dol,Betzios:2018kwn}. Motivated by earlier investigations \cite{Gursoy:2010jh,Gursoy:2010kw}, that found a continuous type Hawking-Page transition in dilaton-gravity with a specific single exponential dilaton potential, these papers studied how the critical behavior arises at late times in a non-conformal version of Bjorken hydrodynamic flow \cite{Gursoy:2015nza} and perturbative fluctuations around this critical limit \cite{Betzios:2017dol,Betzios:2018kwn}. In this theory, criticality arises as a specific limit of the single-exponential dilaton potential, the term that dominates in the IR\footnote{Even though the dilaton becomes large in the IR, the dilaton rescaled by $1/N^2$ remains small hence the loop corrections that would arise when this geometry is embedded in string theory are negligible \cite{Gursoy:2010jh}.}, $V \propto \exp(-8X\phi/3)$ in the limit in which the constant $X$ is sent to $-1/2$. This constant $X$ can be viewed as a conformality breaking parameter in the dual QFT with the conformal theory corresponding to $X=0$.  

An attractive feature of this holographic setting is that a large sector of  the perturbative fluctuations can be solved analytically.  This is because the 5D black hole approaches in the critical limit the product of the 2D linear dilaton black hole solution in string theory \cite{Mandal:1991tz,Elitzur:1990ubs,Witten:1991yr} and $\mathbb{R}^3$. The 2D black hole possesses an $SL(2,\mathbb{R})$ symmetry under which the fluctuations can be organized and the reflection amplitude at the horizon can be expressed analytically in terms of ratios of Gamma functions. In \cite{Betzios:2017dol,Betzios:2018kwn} the quasinormal modes (QNMs) in the $X \to -1/2$ limit were studied and   the energy-momentum tensor correlator was expressed in terms of the reflection amplitude of the 2D string black hole. For fluctuations in the helicity-two sector, two distinct set of special frequencies were found: (i) For  $\textrm{Re}\ \omega/ 2\pi T <   \sqrt{1 + q^2}$, the only would-be QNMs originate from the poles of the reflection amplitude. These poles, however, cannot be extended beyond the near horizon region, hence do not show up as poles of the two-point function. (ii) For $\textrm{Re}\ \omega/ 2\pi T >  \sqrt{1 + q^2}$ there is a different set of QNMs that can indeed  be matched to the poles of the two-point function. 
It was also pointed out in \cite{Betzios:2017dol,Betzios:2018kwn} that one way to define the holographic theory properly, hence {to be} able to match the QNMs with fluctuations of a dual plasma, was to glue the near horizon geometry to $AdS_5$ near the boundary. This procedure yields then a third set of modes that arise from the near boundary AdS region, which we refer to as set (iii). In conclusion, the QNMs  comprise  a set of IR modes that arise from the dilatonic brane (set (ii) above) and a set of UV modes that arise from the AdS region (set (iii))\footnote{This classification holds strictly for fluctuations with helicity two, which we solved analytically in~\cite{Betzios:2018kwn}. The other helicity sectors were analyzed numerically: at finite momentum, the only essential difference is that in addition to the sets (i), (ii), and (iii), these sectors also contain the hydrodynamic diffusion and sound modes.}. 

In connection to approach to criticality, the main result of this analysis was that the non-hydrodynamic quasi-normal modes (QNMs), set (ii), of the black hole coalesce at a branch point on the real axis at $\omega/2\pi T = \sqrt{1 + q^2}$ where $q = k/ 2\pi T$ is momentum in the units of temperature. This means that the hydrodynamic approximation breaks down in the critical 
limit\footnote{See \cite{Arean:2020eus} for a similar conclusion in a different setting whereby criticality is characterized by  $AdS_2$ near horizon regime.}.

Quite interestingly, the critical limit $X \to -1/2$ of 5D dilaton-gravity can be directly related to the large dimension limit in general relativity~\cite{Emparan:2013moa}, see \cite{Emparan:2020inr} for a recent review. This arises because 
the 5D dilaton-gravity with a single-exponential potential 
can be 
obtained from a parent pure AdS theory in D dimensions using generalized dimensional reduction \cite{Gouteraux:2011qh};  then one finds that the extra number of dimensions, hence D, diverges precisely in this critical limit. Another confirmation of this connection with the large D gravity is the fact that the same aforementioned 2D string black hole geometry arises as the near horizon geometry of a generic class of black hole geometries in the large D limit \cite{Emparan:2013xia}. 

A salient feature of the QNMs in the large D limit is that they can be divided into two classes \cite{Emparan:2014cia}, (1) a small number of ``slow modes'' whose frequency do not scale with D, whose spectrum is specific to each geometry, and that are localized in the near-horizon region,  hence these are called the ``decoupled modes''; {they exist only in the helicity-zero and helicity-one sectors; } (2)  the ``fast modes'' whose frequency scales with D and whose spectrum is featureless in the sense that it is determined only by the location and shape of the horizon. These modes extend between the near horizon and the far asymptotic region, hence they are called ``non-decoupled'' modes, {and can be found in all the helicity sectors. 
The universality of the near-horizon region (the 2D black hole) is responsible for the generic behavior of the decoupled modes, whereas the non-decoupled modes (in the asymptotically flat case) arise from a different mechanism: the modes become sensitive only to the peak of the effective radial potential, which becomes universal in the large D limit \cite{Emparan:2020inr}. 

Our analysis is complementary to that of these works. We focus on the helicity-two sector, so we do not see decoupled modes. The set (ii) of IR modes of 5D dilaton-gravity described above can be seen as corresponding to the ``non-decoupled modes'' of the large D limit, since they 
extend from the near-horizon to the asymptotic region. However the shape of the radial potential  is different because of the different  asymptotics (AdS instead of flat, in the D-dimensional theory), and no sharp peak is developed even in the large-D limit.\footnote{See Fig. 1 in \cite{Emparan:2015rva} for an illustration of the potential with different asymptotics.} Despite this, we found that the universality of the near-horizon region leaves an imprint also on this set of QNMs, and is responsible for the critical behavior described above. We will still use the terminology of ``decoupled'' and ``non-decoupled'' modes to denote modes that live in the near-horizon, or extend to the asymptotic region, respectively. 
}

In this paper, we will extend this analysis to a charged plasma that we model by adding a 5D gauge field to dilaton-gravity. We find that {\em charged} critical behavior arises in the same limit in which the geometry again becomes of linear-dilaton type. Now the universal near-horizon geometry of the 2D black hole is replaced by the charged 2D black hole solution \cite{McGuigan:1991qp}. Owing to the emergent SL(2,$\mathbb{R}$) symmetry the (helicity-two) near-horizon fluctuations can also be studied analytically and the poles of the correlation functions in the dual plasma can again be expressed in terms of the reflection amplitude of the charged 2D black hole. Below we summarize our main findings: 
\begin{enumerate}
\item  Interestingly, we find a new set of QNMs that were absent in the neutral solution. These are purely imaginary modes that can be thought of as localized near the horizon, let us denote them as set (iv). Therefore, these are the analog of the ``decoupled'' QNMs in the large D limit. Similarly to set (i), these modes arise from the poles of the reflection amplitude, but unlike set (i), these poles also appear in the full two-point amplitude.

\item These new modes exhibit interesting behavior in the near extremal limit of the charged black hole. In this limit, they become dense at the origin, giving rise to a branch point at $\omega = 0$ thus breaking the hydrodynamic approximation. 

\item  In addition to these new set of modes, there are the trivial extensions of the IR and the UV modes, respectively set (ii) and (iii) above, to the charged case.

\item We study how the QNMs depend on charge and showed that the new modes, set (iv), depend strongly on the charge while the modes in set (ii) show moderate and the ones in set (iii) show very mild dependence of the charge.

\item Finally we develop an alternative WKB method that applies to a more general class of black holes in the critical limit and apply this technique to study the QNM spectrum of the so-called hyperscaling violating Lifshitz black holes \cite{Gouteraux:2011ce}. 

\end{enumerate} 

The organization of the rest of the paper is as follows. In the next section we outline the charged dilatonic black brane solution to 5D Einstein-Maxwell-Dilaton gravity and specify its thermodynamic properties. We show how it arises from higher dimensional AdS 
gravity by compactification and review the connection between the critical limit $X\to -1/2$ and the large D limit of gravity. Finally we point out the relation between the near horizon limit and the 2D charged string black hole \cite{McGuigan:1991qp} together with a review of the scattering amplitude of the 2D black hole. Section~\ref{sec:flucts} contains our main findings where we study the quasinormal mode spectrum of the background. We present analytic expressions for the stress energy correlator and its poles in the critical limit and study how the QNMs depend on charge and temperature. We UV-complete the geometry by gluing an AdS slice near the boundary and discuss how this completion affects the spectrum. Finally, we present an alternative WKB method to compute the quasi-normal modes in the critical limit and employ this to compute the spectrum of hyperscaling violating Lifshitz black holes. We discuss our results with an outlook toward future directions in the last section. 

{Three appendices provide some of the fine points in our computations but also include new material. In appendix A,  we present the details of the UV-IR matching procedure we employ to find the QNM spectrum. Appendix B provides a holographic analysis of the Bjorken solution in the charged plasma. Finally, Appendix C discusses the QNM spectrum in the large momentum limit based on the WKB approximation.}


\section{Charged dilatonic black-brane background} 
\label{sec:chargedCR}
\subsection{The gravitational solution}
Dilaton-gravity in five dimensions has been extensively used in the context of bottom-up holography as an approximation to the strong coupling dynamics of the glue sector of QCD-like gauge theories in the large N limit \cite{Gursoy:2007cb,Gursoy:2007er,Gubser:2008ny}. To use the standard AdS/CFT dictionary between bulk fields and operators one typically considers asymptotically $AdS_5$ geometries. Then the leading term of the dilaton near the boundary is identified with the `t Hooft coupling constant $\lambda$ and the subleading term corresponds to the VeV of the $\tr G^2$ operator. Running of the gauge coupling under the RG flow is then described by non-trivial dependence of the dilaton on the holographic coordinate which can be engineered by choosing the dilaton potential $V(\phi)$. The small/large dilaton asymptotics correspond to small/large 't Hooft coupling, which in turn correspond to the UV/IR limit of the gauge theory with IR slavery. 
The large dilaton asymptotics of the dilaton potential can be chosen such that the horizonless vacuum solution in the 5D theory corresponds to a confining vacuum of the dual gauge theory. At finite temperature there exist black hole (BH) solutions that correspond to the deconfined glue-plasma. There is typically a first order Hawking-Page  transition between the finite $T$ extension of the vacuum solution, i.e. the thermal gas (TG) of gravitons, and the dominant black hole saddle,  which corresponds to a first order confinement-deconfinement transition. 

Fundamental matter has been added to this setup using space-filling flavor branes in \cite{Jarvinen:2011qe} following~\cite{Bigazzi:2005md,Casero:2007ae,Iatrakis:2010zf,Iatrakis:2010jb}. For our purposes, it suffices to consider only a U(1) baryon current which then instructs us to include a 5D gauge field in the bulk. We then consider the following (most general) five-dimensional Einstein-dilaton-Maxwell action\footnote{We use the mostly plus signature for the metric.} 
\begin{equation}\label{action}
 {\cal A} = -\frac{1}{16\pi G_5} \int d^{5}x \sqrt{-g}\left( R - \frac43 (\partial\phi)^2 
+ V(\phi) - \frac{1}{4} Z(\phi) F^2
 \ri)+\,\,\, G.H.\, ,
\end{equation}
%
where ``G.H.'' refers to the Gibbons-Hawking term. In \cite{Gursoy:2007er} it was shown that a confining dual theory requires the large $\phi$ limit of the dilaton potential to be of the form 
\be \label{conf} 
 V(\phi) \to e^{-\frac{8X}{3} \phi} \phi^P \ , 
 \ee
where $X< - 1/2$ with $P$ arbitrary or $X= -1/2$ with $P>0$. The limiting value $X=-1/2$ of this exponent was investigated in \cite{Gursoy:2010jh} in more detail where it was shown to lead to a continuous Hawking-Page transition and the associated critical behavior was further studied in \cite{Gursoy:2010kw}. This particular potential also arises in the context of little string theory and can be embedded in IIB compactifications with NS5 branes to 5D, see e.g. \cite{Aharony:1998ub,Antoniadis:2011qw}.  

Motivated by this, we will consider potentials of single exponential type
\be \label{potentials} 
 V(\phi) = \frac{V_0}{\ell^2} e^{-\frac{8 X}{3}\phi} \ , \qquad Z(\phi) = e^{\frac{8X'}{3}\phi} \ ,
\ee
because, as discussed above, these potentials arise in gravitational duals of confining gauge theories in the IR limit. We absorbed an arbitrary constant in front of the second term in (\ref{potentials}) by a shift in $\phi$ and $\ell$ is a fundamental length scale. In this paper, for simplicity we will focus on the special case $X=X'$ \footnote{We consider the general case $X\neq X'$ in section~\ref{sec:WKB} and in appendix \ref{app:WKB}. These general solutions typically do not correspond to backgrounds with a well-defined holographic dual.}, for which the solution describing a charged black hole in the so-called domain-wall  
coordinates reads: 
\bea \label{metricDW}
 ds^2 &=& \left(\frac{u_0-u}{\ell}\right)^{\frac{1}{2X^2}} \left[ - f(u) dt^2 + \delta_{ij} dx^{i}dx^{j}\right] + f(u)^{-1}d u^2\\
 f(u) &=&  1 - (1+q_0) \left(\frac{u_0-u}{u_0-u_h}\right)^{1-\frac{1}{X^2}} + q_0 \left(\frac{u_0-u}{u_0-u_h}\right)^{-\frac{3}{2X^2}} \\
 \phi(u) &=& \phi_0 +\frac{3}{4X} \log \left(\frac{u_0-u}{\ell}\right) \\
 A_t(u) &=& \mu -  \mu \left(\frac{u_0-u}{u_0-u_h}\right)^{-1-\frac{1}{2X^2}}
 \label{gaugeDW}
\eea
where $u \in (-\infty, u_0)$ is the holographic coordinate and $u_0$ is a curvature singularity that is cloaked by a horizon located at $u_h< u_0$. Regularity of the gauge field at the horizon requires vanishing of $A_t(u_h)$ which can easily be seen in the Euclidean solution where $u_h$ becomes the origin. 
The constants are given as 
\be\lab{intcos} 
\phi_0 = \frac{3}{8X}\log\left(\frac{4V_0 X^4}{3(1-X^2)}\right)\, , \qquad q_0 = \frac{4V_0\mu^2X^4(1+2X^2)}{9(1-X^2)}\left(\frac{u_0-u_h}{\ell}\right)^{2-\frac{1}{2X^2}}\, .
\ee
Here $\mu$ is a dimensionless integration constant that is related to the chemical potential on the boundary as $\mu_{qft} = \mu/\ell$.  

The boundary is located at $u=-\infty$ and $u_0$ is a singularity. This singularity is repulsive, i.e. the bulk fluctuations are suppressed near it in the regime $-\sqrt{2}/2 \leq X \leq -1/2$\footnote{This follows from the analysis of fluctuations around the thermal gas geometry, see \cite{Gursoy:2007er}}. 
The solutions are well defined for $-1<X<0$. However $X = -1/2$ is a special case which corresponds to the borderline case {for the potential} in (\ref{conf}) which should be handled carefully. 

The TG background that corresponds to the ground state of the dual field theory is obtained by requiring vanishing entropy which corresponds to sending $u_h\to u_0$. This background has $A_t(u) = \mu$ and a trivial blackening factor $f=1$. The charge of the ground state vanishes as shown below. 

Another useful set of coordinates is the conformal coordinates that can be obtained from above by
\be \label{relldef}
 r = \frac{4X^2 \ell}{1-4X^2} \le(\frac{u_0-u}{\ell}\ri)^{-\frac{1-4X^2}{4X^2}} \equiv \ell' \le(\frac{u_0-u}{\ell}\ri)^{-\frac{1-4X^2}{4X^2}} \ ,
\ee
for $X\neq -1/2$ and 
\be \label{relldef12}
 \frac{r}{\ell} = - \log\le(\frac{u_0-u}{\ell}\ri) \ ,
\ee
for $X=-1/2$. The boundary is now at $r=0$ for $-1/2<X<0$ and at $r=-\infty$ for $X=-1/2$. The singularity is at $r=\infty$ in both cases. We also defined a new length scale $\ell'$ for later convenience. For $-1<X<-1/2$ the singularity is at $r=0$ and the boundary is at $r=\infty$. The solution for $0>X> -1/2$, which we will be interested in this paper, in these coordinates reads, 
\bea \label{metricCR}
 ds^2 &=&e^{2 A(r)} \left[f(r)^{-1}d r^2 - f(r) dt^2 + \delta_{ij} dx^{i}dx^{j}\right] \\
e^{2A(r)} &=&  \le( \frac{ r}{ \ell'}\ri)^{\frac{2}{3} (-\xi +1)}  \\
  f(r) &=&  1-(1+q_0)\left(\frac{ r}{ r_h}\right)^{ \xi } +q_0 \le(\frac{r}{r_h}\ri)^{2\xi-2}\\
 \phi(r) &=& \phi_0 +\frac{1}{2}\sqrt{(\xi-1)(\xi-4)} \log \left(\frac{r}{\ell'}\right) \\
 A_t(r) &=& \mu-\mu\le(\frac{r}{r_h}\ri)^{\xi-2}
 \label{gaugeCR}
\eea
where we defined the combination
\be \label{xidef}
 \xi \equiv \frac{4(1-X^2)}{1-4X^2} \, , 
\ee
and the constant $q_0$ becomes 
\be\lab{q0r}
q_0 = \frac{(\xi-2)(\xi-4)^2}{18\xi(\xi-1)^2} V_0 \mu^2 \le(\frac{r_h}{\ell'}\ri)^2\, .
\ee
Notice that $\xi \in [4,\infty)$ when $0>X>-{1\over 2}$ and $\xi \to + \infty$ as $X \to -{1 \over 2}$.
One can easily check that one recovers the AdS-Reissner-Nordstrom solution in the limit $X\to 0$ keeping in mind the definition of $\ell'$ in (\ref{relldef}).  
The special solution for $X=-1/2$ in the conformal coordinate system should instead be obtained using the change of variables (\ref{relldef12}) and it reads 
 \bea 
\begin{split}\label{metricCRs}
 ds^2 &=e^{-\frac{2r}{\ell}} \left[f(r)^{-1}d r^2 - f(r) dt^2 + \delta_{ij} dx^{i}dx^{j}\right]\,,\\
 f(r) &= 1-(1+q_0)e^{3\frac{r-r_h}{\ell}} + q_0 e^{6\frac{r-r_h}{\ell}}\, ,\\
 \phi(r) &= \phi_0 + \frac{3r}{2\ell}\,,\\
 A_t(r) &= \mu - \mu\, e^{3\frac{r-r_h}{\ell}}\, ,
 \end{split}
\eea
where 
\be\lab{intcoss} 
\phi_0 =-\frac34 \log (V_0/9) \, , \qquad q_0 = \frac{1}{18} V_0 \mu^2\, .
\ee

To summarise, the solution is determined in terms of two dimensionless parameters in the Lagrangian 
$V_0$ and $X$, which fix  the length scale and determine how the conformality is broken ($X=0$ corresponds to the conformal case) and two dimensionless integration constants $\mu$ and $r_h/\ell$ (or $r_h/\ell'$) which correspond to the chemical potential and the location of the horizon. All other thermodynamic quantities of the dual plasma --- charge, temperature, energy, entropy and pressure --- are determined in terms of these parameters. 

\subsection{Charge, temperature, entropy and mass}

The total charge of the black brane is given by the flux of the electromagnetic potential at the boundary:
\be 
Q_\mathrm{tot}=-\frac{1}{16\pi G_5} \left[ \int\star\left( Z\, F\right)\right]_{boundary} \, ,
\ee
where the dual field strength is given by 
\be
\star F=F_{0 u}g^{00}g^{uu}\sqrt{-g}dx^1\wedge dx^2\wedge dx^3=A_t' e^{A}dx^1\wedge dx^2\wedge dx^3 \ .
\ee
This leads to the following total charge:
\be\label{Qtot}
Q_\mathrm{tot}=\frac{V_3}{24\pi G_5}\frac{\mu V_0 (X^2+2X^4)}{1-X^2}\left(\frac{u_0-u_h}{\ell}\right)^{1+\tfrac{1}{2X^2}} = \frac{3q_0V_3}{32\pi\mu X^2 G_5 }\left(\frac{u_0-u_h}{\ell}\right)^{-1+\tfrac{1}{X^2}} \, .
\ee
The charge of the ground state background follows from (\ref{Qtot}) by sending $u_h\to -\infty$ and it vanishes in the entire allowed range $-1<X\leq 0$.

The temperature of the black brane is obtained by requiring absence of conical singularity on the Euclidean solution as 
\be
  T = e^{A(u_h)}\frac{|f'(u_h)|}{4\pi}\, ,
\ee 
which yields the temperature in terms of $\mu$ and the location of the horizon:  
\begin{align}
\begin{split}\label{thermoT}
  T &=\frac{ 1-X^2}{4\pi \ell X^2} \left(\frac{u_0-u_h}{\ell}\right)^{-1+\tfrac{1}{4X^2}} \left(1- \frac{1+2X^2}{2(1-X^2)}q_0\right)\\
&\equiv \frac{ 1-X^2}{4\pi \ell X^2} \left(\frac{u_0-u_h}{\ell}\right)^{-1+\tfrac{1}{4X^2}} \left(1- \widetilde{Q}^2\right)\, ,
\end{split}
\end{align}
which is a monotonically decreasing function in $u_h$. 
In the last equation we defined the extremality parameter 
\be\lab{extremality}
\widetilde{Q}= \sqrt{2V_0}\mu\frac{(1+2X^2)X^2}{3(1-X^2)}\left(\frac{u_0-u_h}{\ell}\right)^{1-\frac{1}{4X^2}}\, .
\ee
The limit $\widetilde Q \to 1$ corresponds to extremality and we will restrict the parameter space of the solutions to  $0<\widetilde Q \leq 1$. This restriction entails a maximum value for $\mu$ for fixed $u_h$ and for $u_h$ for fixed $\mu$.  In particular, for $X<-1/2$ the limit $u_h\to-\infty$ falls outside of the allowed range unless $\mu$ is simultaneously sent to zero, or $X\to -1/2$. Similarly $u_h \to u_0$ falls outside the allowed range for $X> -1/2$. 

It is easy to invert the expression (\ref{thermoT}) to express the horizon locus as a function of the temperature:
\be\lab{uhT}
\left(\frac{u_0-u_h}{\ell}\right)^{-1+\tfrac{1}{4X^2}}=\frac{2\pi\ell TX^2}{(1-X^2)}\left(1-\sqrt{1+\frac{V_0\mu^2(1+2X^2)^2}{18\ell^2 \pi^2 T^2}}\right) \, .
\ee
The temperature corresponding to the special solution with $X= - 1/2$ reads 
\be
  T = \frac{3}{4\pi\ell}\le(1- q_0\ri) = \frac{3}{4\pi\ell}\le(1-\frac{V_0 \mu^2}{18}\ri) \, .
  \label{thermoTs}
\ee
Interestingly, in the special case the temperature becomes independent of the location of the horizon and the black hole becomes extremal at a specific value of the chemical potential $\mu = 3\sqrt{2/V_0}$. The temperature of the TG solution is a free parameter which we set equal to the temperature of the black hole.

The entropy   of the black hole can be computed as usual as 
\be\lab{ents}
S_{\mathrm{tot}}=\frac{V_3}{4 G_5} \left(\frac{u_0-u_h}{\ell}\right)^{\tfrac{3}{4X^2}}\, .
\ee
Entropy of the TG solution is suppressed as $1/N^2$ (equivalently by the Newton's constant $G_5$) in the large $N$ limit, therefore it vanishes in the leading order.

The ADM mass of the solution can be obtained by first considering a time slicing of the geometry:
\be
ds^2 = -N^2 dt^2 +\gamma_{mn}\left(dx^m -N^m dt\right)\left(dx^n -N^n dt\right)\, , \qquad \qquad  m,n= u,1,2,3\, ,
\ee
where $\gamma_{mn}$ is the induced metric on the 4D slice of constant time. Then the mass is given by \cite{Gursoy:2008za}:
\be
E_\mathrm{tot} = -\frac{1}{8\pi G_5} \int N\left(\sqrt{\gamma} \, K - \sqrt{\gamma_0}\, K_0 \right)\, ,
\ee
where $K$ is the 3D extrinsic curvature of the geometry, quantities with a subscript $0$ refer to the thermal gas solution which we introduced below  \eqref{intcos} and the integral is computed at spatial infinity for constant time. Substituting the solution \eqref{metricDW} we find
\be \label{ADM}
E_\mathrm{tot} = \frac{3V_3}{2\pi G_5}\lim_{u_c\rightarrow -\infty} e^{4A(u_c)} \sqrt{f(u_c)} A'(u_c) \left(\sqrt{f(u_c)}-1\right) = \frac{3 V_3(1+q_0)}{64\pi \ell X^2 G_5,} \left(\frac{u_0-u_h}{\ell}\right)^{-1+\tfrac{1}{X^2}}\, ,
\ee
where we used that $A_0(u) = A(u)$ and $ f_0(u)=1$. 
In the extremal case we have the following relation between energy and charge of the black hole:
\be
E_\mathrm{tot} = \frac{3}{4(1-X^2)} \frac{\mu Q_{\mathrm{tot}}}{\ell}\,.
\ee
The energy of the special solution $X=-1/2$ reduces to
\be 
E_\mathrm{tot} = \frac{3(1+q_0)}{4\ell \pi}S_{\mathrm{tot}} = \frac{18+\mu V_0^2}{24\ell \pi}S_{\mathrm{tot}}\, ,
\ee
i.e. the $X=-1/2$ solution has an entropy that is linear in energy $S \propto E$, that is interestingly the typical Hagedorn behavior in string theory, see \cite{Chen:2021dsw} for a recent discussion. It also agrees with the large D limit of black holes \cite{Emparan:2013xia}.

\subsection{Thermodynamics} \label{Thermo1}

The free energy of the plasma --- that is the same as the pressure in an extensive system --- is  obtained from the gravitational action evaluated on-shell 
\be
\Omega = \frac{\Delta\mathcal{A}_{on}}{\beta}\, ,
\ee
where $\Delta \mathcal{A}_{on}$ is the (regularized) on-shell action and $\beta$ the inverse temperature. The action is infinite due to divergences near the boundary. This can be regularized by standard holographic renormalization. However, since we are typically interested in the differences of energies between the thermodynamic states here we  simply cut off the radial direction\footnote{ In general case, one needs to define two different UV cut-off surfaces $u_c$ and $u_c^0$ for the BH and TG solutions  respectively which are determined in the UV by a set of consistency conditions~\cite{Gursoy:2008za}. In our case, however, the TG solution is obtained from the solution in~\eqref{metricCR}--\eqref{gaugeCR} simply by taking $u_h \to -\infty$ and the $u_h$ only appears in the blackening factor; in particular the dilaton solution is independent of $u_h$. This implies $u_c=u_c^0$.} at $u=u_c$ and subtract there the on-shell actions of the BH and the TG solutions with same values for the external quantities i.e. the same $T$ and $\mu$. 

The regularized on-shell action for the solution \eqref{action} is then found by adapting the calculation in \cite{Gursoy:2008za} to the charged black holes \cite{Gursoy:unpublished} which in general contain a nontrivial blackening factor also for the TG geometry 
\be 	\label{Son1}
 \frac{\Delta \mathcal{A}_{on}}{\beta}= \frac{ V_3 e^{4A(u_c)}}{16\pi G_5 }\left[f'(u_c)-f'_0(u_c)\sqrt{\frac{f(u_c)}{f_0(u_c)}}+6\left(A'(u_c)f(u_c)-A'_0(u_c)\sqrt{f(u_c)f_0(u_c)}\right)\right] \ , 
\ee
where $A_0$ and $f_0$ are the conformal and the blackening factors corresponding to the TG solution. For the solution we consider here, $f_0=1$ and $A_0(u)=A(u)$. 

Using the near boundary expansions of the background functions it is straightforward to show that the first term in (\ref{Son1}) gives the sum of enthalpy and the electric energy difference between the BH and the TG solutions,  $-T S_\mathrm{tot}-\mu Q_\mathrm{tot}/ \ell$,  and the second vanishes in leading order in N. A similar calculation shows that the last term matches precisely the ADM mass (\ref{ADM}) which is the defined as the difference between the black hole and thermal gas ADM integrals. One finds the following analytic expression in $T$ and $\mu$ for the free energy in the grand canonical ensemble: 
\be\lab{Smarr} 
\Omega = -\frac{V_3}{64\pi \ell G_5}\le( c_1 \left(\frac{u_0-u_h}{\ell}\right)^{-1+\tfrac{1}{X^2} } + \mu^2 c_2 \left(\frac{u_0-u_h}{\ell}\right)^{1+\tfrac{1}{2X^2} }  \ri), 
\ee
where $(u_0-u_h)$ is expressed in terms of $T$ and $\mu$ in equation (\ref{uhT}) and the coefficients $c_1$ and $c_2$ are given by 
\be\lab{c1c2}
c_1 = \frac{1-4X^2}{X^2}\, , \qquad c_2 = \frac{4V_0 X^2(1+2X^2)(1-4X^2)}{9(1-X^2)}\, .
\ee 
It is straightforward to check that this satisfies the Smarr relation:
\be\lab{Smarr1p5} 
\Omega = E_\mathrm{tot} -T S_\mathrm{tot}-\frac{\mu}{\ell} Q_\mathrm{tot} \, .
\ee
We also observe that the energy of the system in this plasma is proportional to the sum of the enthalpy and the electric energy, therefore the free energy becomes
\be\lab{Smarr2} 
\Omega = -\frac{1}{\xi}\le[T S_\mathrm{tot}+\frac{\mu}{\ell} Q_\mathrm{tot}\ri] = - \frac{1}{\xi-1} E_\mathrm{tot}  \, ,
\ee
where $E_\mathrm{tot}$ is given in (\ref{ADM}). The final expression for the free energy in terms of the physical parameters $\mu$ and $T$ follows from (\ref{uhT}) and it is straightforward to check that 
\be
\frac{\partial\Omega}{\partial T}=-S_\mathrm{tot}\, ,\qquad\frac{\partial\Omega}{\partial \mu}=-Q_\mathrm{tot}\, .
\ee
It is clear from the expressions above that the free energy has a definite sign 
and generically does not vanish. Therefore there is generically no phase transition in this theory between the ground state and the plasma phase. For $0>X > -1/2$ the plasma phase always lowers the free energy hence dominates the grand canonical ensemble; the opposite happens for $-1 < X < -1/2$.  For $X= -1/2$ the free energy vanishes identically confirming our expectation that the $X = -1/2$ limit focuses on the critical point. The special locus $u_h = u_0$, that is when the (infinitesimally small) black hole marginally cloaks the singularity, should be analyzed separately. This limit is never reached for $0>X > -1/2$ for positive temperature; see the discussion below equation (\ref{extremality}). On the other hand, it corresponds to $T\to \infty$ for $-1/2 > X > -1$.  

%
%
\subsection{Dimensional reduction and the connection to large $D$}

As it was shown in \cite{Gursoy:2015nza} for the case without Maxwell field, the action~\eqref{action} with the exponential dilaton potential  ~\eqref{potentials} can be obtained formally from a generalized dimensional reduction of Einstein gravity with a cosmological term. We repeat the derivation here, including also the gauge field. We start with the Einstein-Maxwell action in 
$\xi +1$ dimensions, 
\begin{equation} \label{dimregaction}
 {\cal A} = \frac{1}{16\pi \tilde G_N} \int d^{d+1}x d^{\xi-d} y \sqrt{-\tilde g}\left( \tilde R - 2 \Lambda - \frac{1}{4} F^2 
 \ri) \,, 
\end{equation}
and we take an Ansatz for the metric of the form 
\be \label{dimregansatz}
\widetilde{ds}^2 = e^{-\delta_1 \phi(x)} g_{\mu\nu}dx^\mu dx^\nu + e^{\delta_2 \phi(x) } dy^2 \, 
\ee
where the metric in the $x$ directions is generic, while in the $y$ directions it is taken to be flat (so we compactify on a torus $T^{\xi-d}$).
Requiring the reduced action to be in Einstein frame, and the dilaton potential, which comes from the cosmological constant term, to be 
$e^{- {2d \over d-1} X \phi}$, gives 
\be 
\delta_1 = {2d \over d-1} X \,, \quad \delta_2 = {2d \over \xi -d} X \,.
\ee
We further require that the dilaton is normalized with a prefactor ${d\over d-1}$ (which corresponds to a canonically normalized dilaton in the string frame); this fixes the value of the number of extra dimensions:
\be 
\xi = {d (1-X^2)\over 1 - d X^2} \,, 
\ee
which coincides with~\eqref{xidef} for $d=4$. 
Finally, the Maxwell term reduces to the lower-dimensional Maxwell action with 
\be 
Z(\phi) = e^{ {2 d \over d-1}X \phi} 
\ee 
which corresponds to \eqref{potentials} with $X'=X$. 

The previous calculation shows that the critical limit $\xi \to \infty$ can be equivalently interpreted as a large $D$ limit, where $D=\xi+1$. 
The large D expansion of gravity, pioneered in \cite{Emparan:2013moa,Emparan:2013xia}, has been recently applied in many situations as a way of simplifying Einstein equations 
and obtaining analytic results, {\it e.g.} on the quasi-normal spectrum of black branes \cite{Emparan:2014cia,Emparan:2015rva}. 

We consider the dimensional reduction of time-dependent solutions and the Bjorken flow in Appendix~\ref{app:Bjorken}.

\subsection{Connection to 2D black holes }

Neutral black holes exhibit universal near horizon behavior in the large $D$ limit as shown in \cite{Emparan:2013xia}. The near horizon geometry shared by these black holes is the two-dimensional neutral black hole of \cite{Dijkgraaf:1991ba} which is supported by a linear dilaton. As argued in the previous section, the parameter $\xi$ in our model, when the model is viewed as dimensionally reduced from a parent theory, is related to  the number of extra dimensions in this parent theory. Therefore it is reasonable to expect that the near horizon geometry of the large $\xi$ solution  in \eqref{metricCRs} becomes the two-dimensional \textit{charged} analogue of \cite{Dijkgraaf:1991ba}, i.e. the black hole studied in \cite{McGuigan:1991qp}. To demonstrate that this is indeed the case, we first switch to string frame:
\be
ds^2_{string}=e^{\frac{4}{3}\phi(r)}ds^2_{Einstein}\, .
\ee
Then the solution~\eqref{metricCRs} becomes
\begin{align}
\begin{split}
 ds^2 &=e^{\frac{4\phi_0}{3}} \left[f(r)^{-1}d r^2 - f(r) dt^2 + \delta_{ij} dx^{i}dx^{j}\right]\, ,\\
 f(r) &= 1-(1+q_0)e^{3\frac{r-r_h}{\ell}} + q_0 e^{6\frac{r-r_h}{\ell}}\, ,\\
 \phi(r) &= \phi_0 + \frac{3r}{2\ell}\, ,\\
 A_t(r) &= \mu - \mu\, e^{3\frac{r-r_h}{\ell}}\, ,
\end{split}
\end{align}
where $q_0$ is given in (\ref{intcoss}).  The two-dimensional part of the background --- obtained by reducing on the transverse directions $x^i$ --- can be rewritten as follows by a coordinate transformation 
\begin{align}
\begin{split}  \label{metric2D}
\frac{4}{k}ds^2_{2D}&=-dt^2 f(\rho)+\frac{\ell^2 d\rho^2}{9\rho^2f(\rho)}\,,\\
 f(\rho)&=1-\frac{2M}{\rho}+\frac{p^2}{\rho^2}\,, \\
 \phi(\rho) &= \phi_0 - \frac{1}{2}\log \frac{3\rho}{\ell}\,, \\
 A_t(\rho) &= \mu \left(\frac{1}{\rho_+}-\frac{1}{\rho} \right)\, ,
\end{split}
\end{align}
where we make the identifications 
\begin{align}
\begin{split}
\frac{4}{k}&\equiv e^{-\frac{4 \phi_0}{3}} \label{2dk}  \, , \\
 \rho&\equiv e^{-3\tfrac{r-r_h}{\ell}}  \, .
\end{split}
\end{align}
Mass and charge of the 2D black hole are 
\be
M=\frac{1+q_0}{2}\, ,~~~~~~~~~~~~~~~~~~~~~~~~ p^2= q_0\, ,
\ee
and the two horizons are located at 
\begin{align}
\begin{split}
\rho_+& =  M+\sqrt{M^2-p^2} = 1\, ,\\
\rho_- & = M-\sqrt{M^2-p^2}= q_0\, ,
\end{split}
\end{align}
where we recall that $q_0 \in [0,1] $.
Propagation of strings in this geometry can be described by the coset model $\frac{SL\le(2,\mathbb{R}\ri)_k \times U(1)}{U(1)}$ \cite{Giveon:2005jv} in which the constant $k$ introduced in \eqref{2dk} plays the role of the level in the $SL\le(2,\mathbb{R}\ri)$ algebra. 
 
It is possible to compute the reflection amplitude for a wave scattered off the event horizon of this 2D black hole. The result is given by \cite{Elitzur:2002rt}
\be
R=\frac{\Gamma\le(1+i \frac{s}{k}\ri)}{\Gamma\le(1-i \frac{s}{k}\ri)}\frac{\Gamma\le(i s\ri)}{\Gamma\le(-i s\ri)}\frac{\Gamma\le(\frac{1}{2}\le(1-is+i m\ri)\ri)}{\Gamma\le(\frac{1}{2}\le(1+is+i m\ri)\ri)}\frac{\Gamma\le(\frac{1}{2}\le(1-is-i \bar{m}\ri)\ri)}{\Gamma\le(\frac{1}{2}\le(1+is -i \bar{m}\ri)\ri)},
\ee
where the incoming wave has energy $\propto(m-\bar{m})$ and momentum $s\in \mathbb{R}$.  Assuming the incoming wave is neutral, $m$ and $\bar{m}$ are related to each other by the gauge condition \cite{Giveon:2003ge}
\be
{m}=-\frac{1+p^2}{1-p^2}\bar{m}.
\ee
Moreover one can take the so called \textit{minisuperspace} or semiclassical approximation\footnote{Equations (\ref{2dk}) and (\ref{intcoss}) seem to relate the minisuperspace approximation to the limit $V_0\to 0$. This is in fact not the case as an additive integration constant of the dilaton, which we omitted above, enters the relation between the string coupling constant and the dilaton.  This should scale as $\log N$ to match the 't Hooft coupling in the dual theory. Thus, large $k$ means large $N$ in the field theory and this limit should be assumed in order to omit the loop corrections in the string theory. See \cite{Gursoy:2010jh} for a detailed discussion.}, i.e. $k\rightarrow \infty$, such that the reflection amplitude becomes
\be\lab{R2D}
R=\frac{\Gamma\le(i s\ri)}{\Gamma\le(-i s\ri)}\frac{\Gamma\le(\frac{1}{2}\le(1-is-\frac{1+p^2}{1-p^2}i \bar{m}\ri)\ri)}{\Gamma\le(\frac{1}{2}\le(1+is-\frac{1+p^2}{1-p^2}i \bar{m}\ri)\ri)}\frac{\Gamma\le(\frac{1}{2}\le(1-is-i\bar{m}\ri)\ri)}{\Gamma\le(\frac{1}{2}\le(1+is -i \bar{m}\ri)\ri)}.
\ee
We note that setting $p=0$ correctly gives the reflection amplitude for the $2D$ uncharged black hole \cite{Dijkgraaf:1991ba}.

We will match this reflection amplitude with the one we compute by solving the fluctuation equation near the horizon in the next section where we will show that $m$ is determined in terms of the frequency of the wave, $s$ is determined by frequency and momentum and $p$ corresponds to the extremality parameter $\tilde Q$. 

\section{Analytic results for the fluctuations around the charged dilatonic black hole} \label{sec:flucts}

\subsection{Solution of the fluctuation equations} \label{sec:fluctsol}

In this section we study the fluctuations around the background~\eqref{metricCR}, 
which we will refer to as the Chamblin-Reall (CR) background below~\cite{Chamblin:1999ya}, and find its quasi-normal modes. These quasi-normal modes, apart from their important role in the study of black holes \cite{Berti:2009kk} correspond to poles of the retarded Green's functions in the plasma phase of the corresponding dual gauge theory and determine how small fluctuations equilibrate at late times. 
We will concentrate on the helicity-two fluctuations and comment on the other sectors briefly below. 
In the helicity-one and helicity-zero sectors various modes mix, in general, 
which complicates the analysis.
Finite momentum fluctuations take the same form as the vanishing momentum, hence we consider the general case with momentum which we take to be in the $x_1$ direction. We write for the fluctuation of the metric
\be
 \delta g = e^{2 A(r)} H(r)e^{-i\omega t+ikx_1} dx_2 dx_3 \ .
\ee
The resulting fluctuation equation can be written in conformal coordinates as
\be \label{h2eq}
\left(\frac{\omega ^2}{f(r)}-k^2\right) H(r) + \frac{f(r)-\xi +\xi \widetilde Q^2 \left(\frac{r}{r_h}\right)^{2(\xi-1)} }{r} H'(r) +f(r) H''(r) = 0 \ ,
\ee
{
where the blackening factor simplifies to
\be
 f(r) = 1 -\left(\frac{r}{r_h}\right)^\xi - \frac{\xi\widetilde Q^2}{\xi-2}\left(\left(\frac{r}{r_h}\right)^\xi-\left(\frac{r}{r_h}\right)^{2\xi-2}\right)\ .
\ee
The fluctuation equation} coincides with the equation of a massless scalar field in the charged CR background. As the function $f$ only depends on the ratio $r/r_h$, see \eqref{metricCR}, one can easily factor out $r_h$ from this equation by changing $\omega \to \omega/r_h$, $k\to k/r_h$ and $r \to r r_h$. 

\begin{figure}[t]
\centering
\includegraphics[width=0.5\textwidth]{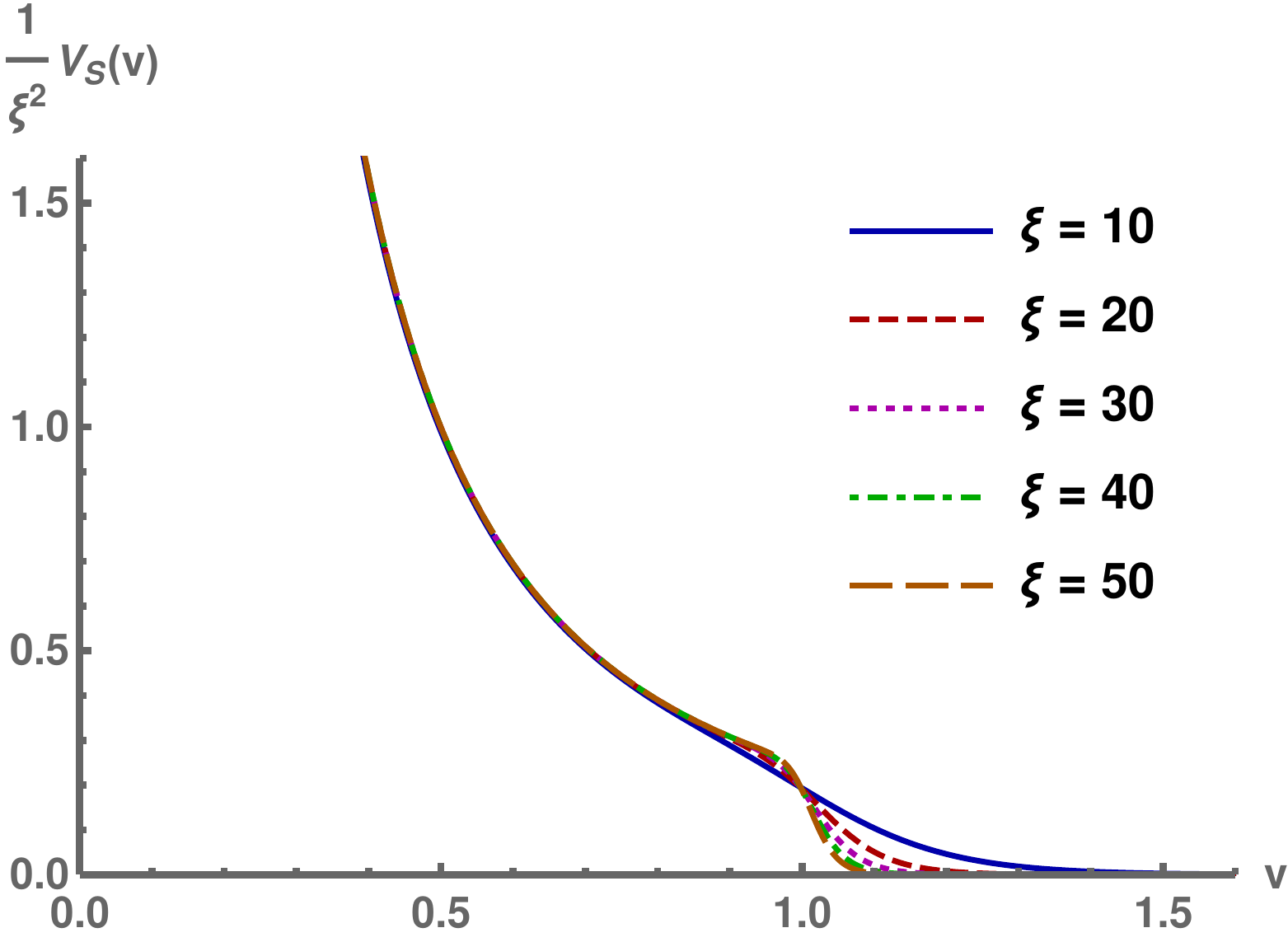}%
\includegraphics[width=0.5\textwidth]{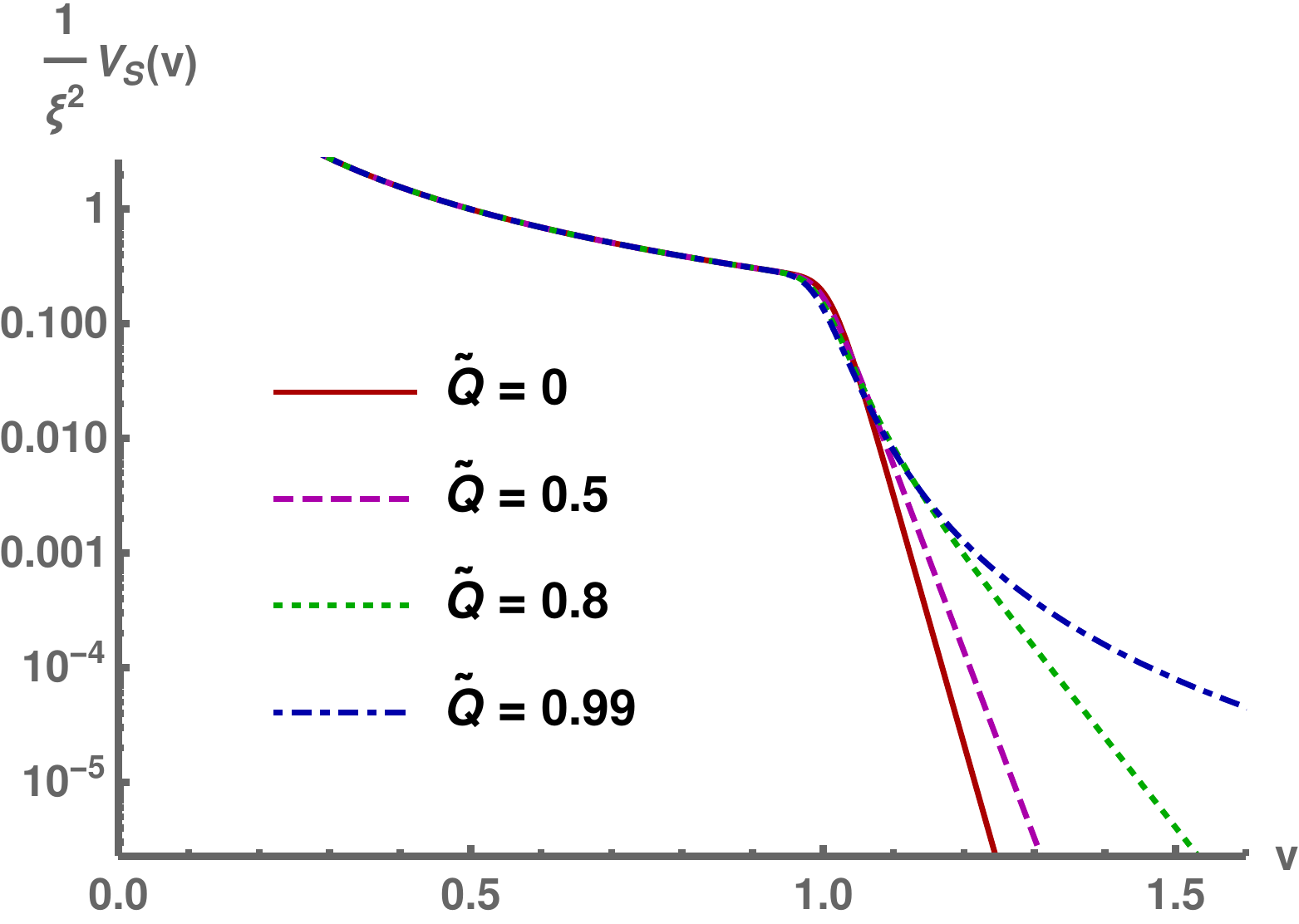}
\caption{
Left: The dependence of the Schr\"odinger potential of the helicity-two fluctuations on $\xi$ at zero charge and momentum. Right: The dependence of the potential on charge at zero momentum and at $\xi=50$.
}
\label{fig:VS}
\end{figure}

{The Schr\"odinger form of the fluctuation equation is obtained by  defining the Schr\"odinger wave function as $\psi(r) = r^{(1-\xi)/2}H(r)$ and by switching to the Schr\"odinger coordinate $v$ defined through
\be
 v'(r) = \frac{1}{f(r)} \ .
\ee
At nonzero charge the relation between $v$ and $r$ cannot be solved analytically. We choose $v(r=0)=0$, so $v$ ranges from zero to infinity as $r$ varies form zero to $r_h$. Near the boundary, $v \approx r$. 
The fluctuation equation for $\psi$ reads
\be \label{eq:Sform}
 \psi''(v) +\left(\omega^2 - V_S(q,v)\right)\psi(v) = 0 \ .
\ee
The Schr\"odinger potential $V_S$ cannot be solved analytically in terms of $v$, but it admits a rather simple expression in terms of the conformal coordinate $r$:
\be \label{eq:VSres1}
 V_S(q,r) = V_{S0}(r) + q^2 f(r) \ ,
\ee
where
\be \label{eq:VSres2}
V_{S0}(r) = \frac{(\xi-1)f(r)}{4r^2}\left(2\xi-(\xi-1)f(r)-2 \xi \widetilde Q^2\left(\frac{r}{r_h}\right)^{2\xi-2} \right) \ .
\ee

We show the dependence of the Schr\"odinger potential on $\xi$ in Fig.~\ref{fig:VS} (left), at zero charge and momentum. Since $r_h$ scales out of the fluctuation equation, we were able to set it to one without loss of generality. We divided out the quadratic $\xi$-dependence near the boundary, where $V_S \sim \xi^2/(4v^2)$. The only nontrivial feature of the potential is the step near $v=1$ which becomes steeper with increasing $\xi$. This steps arises from the factor of $f(r)$ which multiplies the whole potential as we see from~\eqref{eq:VSres1} and~\eqref{eq:VSres2}. The decay at large $v$ is exponential and becomes steeper as $\xi$ increases. The charge dependence is shown in Fig.~\ref{fig:VS} (right). The dependence is mild, and mostly observed in the asymptotic decay at large $v$. To show this effect, we have chosen to plot the potential in logarithmic scale. The dependence of the potential on the momentum $q$ (not shown) is noticeable when the momentum is large with respect to $\xi$, i.e., $q \gtrsim \xi$. Increasing the momentum enhances the normalization of the potential at large $v$ without changing the qualitative behavior. 

We therefore conclude that the Schr\"odinger potential is a relatively simple and featureless function for our setup. However, as we shall see below, the spectrum of QNMs has an interesting structure despite the simplicity of the potential.
} 

As anticipated above, the fluctuations can be solved at large $\xi$ by considering the fluctuations in the UV and near the horizon separately and by matching them in the middle.
The UV solution may be obtained by taking $\xi \to \infty$ with fixed $r<r_h$, $\widetilde Q$, $\omega$, and $k$, which amounts to setting $f=1$ and $\widetilde Q=0$ in~\eqref{h2eq}. As the result is independent of $\widetilde Q$ it coincides with the neutral case~\cite{Betzios:2018kwn}:
\be \label{bdrysol}
 H(r) = C_1 r^{\xi /2} J_{\frac{\xi }{2}}\left(r \sqrt{\omega^2-k^2}\right) +  C_2 r^{\xi /2} J_{-\frac{\xi }{2}}\left(r \sqrt{\omega^2-k^2}\right) \ .
\ee

In order to study the near-horizon limit, we define the rescaled variables 
\be \label{varpidef}
\varpi = \frac{2 \omega r_h} {\xi} =  \frac{\omega}{2\pi T}\left(1-\widetilde Q^2\right) , \qquad q = \frac{2 k r_h} {\xi}=  \frac{k}{2\pi T}\left(1-\widetilde Q^2\right)
\ee
which equal, at zero charge, the ``standard'' normalized frequency and momentum, $\omega/2\pi T$ and $k/2\pi T$, respectively. Moreover we define a new radial coordinate
\be
 w = \le(\frac{r}{r_h}\ri)^\xi \ ,
\ee
and take $\xi \to \infty$ keeping $\varpi,q$ and $w$ fixed. The result reads
\bea
\begin{split} \label{NHflucteq}
 (1-w) \left(1-\widetilde Q^2 w\right) H''(w)-\left(1-\widetilde Q^2 (2 w-1)\right) H'(w) & \\
  +\frac{ \left(\frac{\varpi ^2}{(1-w) \left(1-\widetilde Q^2 w\right)}-q^2\right)}{4 w^2}H(w) &= 0 \ ,
\end{split}
\eea
which is solved by
\begin{align} \label{1F2sols}
H(w) &= C_1 w^{\frac{1}{2}+\frac{1}{2} i \widetilde S} \left(1-\widetilde Q^2 w\right)^{-\frac{1}{2}+\frac{-1+\widetilde Q^2 (1+i \varpi )}{2 \left(\widetilde Q^2-1\right)}} (1-w)^{-\frac{1}{2}+\frac{1}{2} i \left(-\widetilde S+\frac{\widetilde Q^2 \varpi }{1-\widetilde Q^2}\right)} & \nonumber\\
&\, _2F_1\left(\frac{1}{2} \left(1-\frac{1+\widetilde Q^2}{1-\widetilde Q^2}i\varpi+i \widetilde S\right),\frac{1}{2} \left(1+i \varpi + i\widetilde S\right);1+ i \widetilde S;\frac{\left(1-\widetilde Q^2\right) w}{w-1}\right)& \\
\label{1F2sols2}
&+C_2 w^{\frac{1}{2}-\frac{1}{2} i \widetilde S} \left(1-\widetilde Q^2 w\right)^{-\frac{1}{2}+\frac{-1+\widetilde Q^2 (1+i \varpi )}{2 \left(\widetilde Q^2-1\right)}} (1-w)^{-\frac{1}{2}+\frac{1}{2} i \left(\widetilde S+\frac{\widetilde Q^2 \varpi }{1-\widetilde Q^2}\right)}&\nonumber\\
&\, _2F_1\left(\frac{1}{2} \left(1-\frac{1+\widetilde Q^2}{1-\widetilde Q^2}i\varpi-i \widetilde S\right),\frac{1}{2} \left(1+i \varpi - i\widetilde S\right);1+ i \widetilde S;\frac{\left(1-\widetilde Q^2\right) w}{w-1}\right) &
\end{align}
where we defined 
\be\lab{tildeS} 
\widetilde S = \sqrt{\varpi^2 - q^2 -1}\, .
\ee
The regularity at horizon implies
\bea
\begin{split} \label{Refdef}
 \frac{C_2}{C_1} \equiv \mathcal{R} &= -\left(1-\widetilde Q^2\right)^{-i \widetilde S}\\
 &\times \frac{ \Gamma \left(1+i \widetilde S\right) \Gamma \left(\frac{1}{2} \left(1-i \varpi -i \widetilde S\right)\right)  \Gamma \left(\frac{1}{2} \left(1-\frac{1+\widetilde Q^2}{1-\widetilde Q^2}i\varpi-i \widetilde S\right)\right)}{\Gamma \left(1-i \widetilde S\right) \Gamma \left(\frac{1}{2} \left(1-i \varpi +i \widetilde S\right)\right) \Gamma \left(\frac{1}{2} \left(1-\frac{1+\widetilde Q^2}{1-\widetilde Q^2}i\varpi+i \widetilde S\right)\right)} \ .
\end{split}
\eea
The dependence on the charge in the correlator only appears through this modified reflection amplitude. 

Some comments are in order. First, this result matches perfectly the reflection amplitude of the charged dilatonic 2D black hole geometry in the semiclassical limit, which we reviewed in the previous section, see equation (\ref{R2D}). To see this one should make the identifications
\be
s=\widetilde{S}, ~~~~~~~~~~~~~~~~~~~ \bar{m}=\varpi, ~~~~~~~~~~~~~~~~~~~  p=\widetilde{Q} \ ,
\ee
with  $\varpi$ being the rescaled frequency defined in (\ref{varpidef}).  It is also possible to get rid of the prefactor, $(1-\widetilde{Q}^2)^{-i \tilde{S}}$,  in front of the $5D$ reflection amplitude above by making a change of variable (e.g. $\frac{1-\tilde{Q}^2}{1-w}w=\sinh^{-2} \theta$). Notice that this factor is a phase in the region of validity of the coset description, since $\widetilde{S}$ has to be real. The same reasoning can be applied to the reflection amplitude in the extremal case.

Second, we note that the reflection amplitude has poles which were absent at zero charge. They are obtained by solving
\be
 \frac{1}{2} \left(1-\frac{1+\widetilde Q^2}{1-\widetilde Q^2}i\varpi+  S \right) = 1 -n 
\ee
where $S = -i \widetilde{S}$ and $n=1,2,3,\ldots$. The solutions are given by
\be \label{immmodesCR}
 \varpi = -i\left[\frac{\kappa_Q (2 n-1) + \sqrt{(2 n-1)^2+\left(\kappa_Q^2-1\right) \left(q^2+1\right)}}{\kappa_Q^2-1}\right]
\ee
where $\kappa_Q = (1+\widetilde Q^2)/(1-\widetilde Q^2)$
(the expression with minus sign in front of the square root does not solve the equation). In the extremal limit $\widetilde Q \to 1$ the poles become dense:
\be
 \varpi = -i  \left(2 n-1+\sqrt{q^2+1}\right)(1-\widetilde Q) + \mathcal{O}\left(\left(1-\widetilde Q\right)^2\right) \ .
\ee
Moreover the poles and zeros of the reflection amplitude merge into a branch cut in the extremal limit:
\be \label{imbranchcut}
\left(1-\widetilde Q^2\right)^{S} \frac{\Gamma \left(\frac{1}{2} \left(1-\frac{1+\widetilde Q^2}{1-\widetilde Q^2}i\varpi + S\right)\right)}{\Gamma \left(\frac{1}{2} \left(1-\frac{1+\widetilde Q^2}{1-\widetilde Q^2}i\varpi - S\right)\right)} = (-i \varpi)^{S} +   \mathcal{O}\left(1-\widetilde Q\right) \ .
\ee


In passing, let us briefly discuss the other helicity modes. 
The fluctuations for other sectors are complicated for generic $q$, and we do not discuss them here. At  $q=0$, however, there are drastic simplifications. In the absence of preferred direction given by the momentum, all spin two fluctuations must also satisfy Eq.~\eqref{h2eq} with $q=0$, which can be checked explicitly. As it turns out, also the spin zero fluctuation (i.e. the fluctuation of the dilaton) satisfies exactly the same equation. 

The gauge field fluctuations, however, obey a slightly different radial fluctuation equation. It is given by 
\begin{align}
\begin{split}
 &\left(\frac{\omega ^2}{f(r)}- \frac{2\xi (\xi -1) \widetilde Q^2  \left(\frac{r}{r_h}\right)^{2( \xi -1)}}{r^2} \right) a(r)\\
 &+ \frac{3 f(r)-\xi + \xi \widetilde Q^2 \left(\frac{r}{r_h}\right)^{2(\xi-1)} }{r} a'(r) +f(r) a''(r) = 0 
\end{split}
\end{align}
and therefore contains an extra term with respect to the helicity two equation. In the limit of large $\xi$, at fixed $w$, we obtain
\bea
\begin{split}
 (1-w) \left(1-\widetilde Q^2 w\right) a''(w)-\left(1-\widetilde Q^2 (2 w-1)\right) a'(w) & \\
  + \left(\frac{\varpi ^2}{4 w^2(1-w) \left(1-\widetilde Q^2 w\right)} - 2 \widetilde Q^2\right)a(w) &= 0 \ .
  \end{split}
\eea
Because of the extra term this equation cannot be reduced to the hypergeometric equation. (After a change of variables, it reduces to the Heun equation.) Consequently, we have not been able to find the solution and the reflection amplitude for the gauge fields analytically.

%

\subsection{Analytic correlators of the energy-momentum tensor at large $\xi$}\label{sec:ancorrs}

Because at large $\xi$ the charge only modifies the geometries near the horizon, it is straightforward to compute the correlators of the transverse components of the energy momentum tensor at finite charge following the steps given in~\cite{Betzios:2018kwn}. When combined, the near boundary solutions~\eqref{bdrysol} and near horizon solutions~\eqref{1F2sols}--\eqref{1F2sols2} provide complete control over the solution in the limit of large $\xi$. The relations between the integration coefficients are obtained by matching the solutions in the middle where their regimes of validity overlap. 

In the expressions~\eqref{1F2sols}--\eqref{1F2sols2} the dependence on $\widetilde Q$ only appears in subleading terms of the UV expansion at $w=0$ for each of the functions. Since only the leading terms are needed when matching with the UV solutions (which are likewise independent of the charge) the matching procedure is unchanged with respect to the neutral computation. Notice however that the definition of the variable $w$, for example, is the same as in~\cite{Betzios:2018kwn} when expressed in terms of $r_h$ but is changed if expressed through the temperature in~\eqref{thermoT}.  Similar comments apply to the definitions of the rescaled frequency $\varpi$ and momentum $q$ in~\eqref{varpidef}. That is, the effect of the charge is taken into account by substituting the expression~\eqref{Refdef} for the reflection amplitude in the results of~\cite{Betzios:2018kwn}, and recalling that the relation between the temperature and the model parameters is modified. We will review here the main points, and write down the results with charge included. Many of the results will also be rederived in Sec.~\ref{sec:WKB} by using a slightly different approach.

\subsubsection{Charged CR geometry}

The transverse correlator for the charged CR geometry is obtained as in Sec.~4.1.2 of~\cite{Betzios:2018kwn}. It is given by
\bea 
\begin{split}\label{corrsmallomega}
  G_\mathrm{reg} &= \frac{2\pi\, \xi^\xi r_h^{-\xi}}{\Gamma\left(\frac{\xi}{2}\right)\Gamma\left(1+\frac{\xi}{2}\right)} \left(\frac{\left(\varpi ^2-q^2\right)}{16}\right)^\frac{\xi}{2}
  \left[\left(\frac{1+\Sq}{1-\Sq}\right)^\frac{\xi}{2}\, e^{-\xi \Sq}\, \mathcal{R}- i\theta(-\mathrm{Im}\,\varpi) \right]\\
  &=- \frac{2\pi\, \xi^\xi r_h^{-\xi}\, e^{-\xi \Sq}}{\Gamma\left(\frac{\xi}{2}\right)\Gamma\left(1+\frac{\xi}{2}\right)}\left(\frac{\left(\varpi ^2-q^2\right)}{16}\right)^\frac{\xi}{2} \left(\frac{1+\Sq}{1-\Sq}\right)^\frac{\xi}{2}\\ 
  &\quad \times  \left(1-\widetilde Q^2\right)^S\frac{\Gamma \left(1-\Sq\right)}{\Gamma \left(1+\Sq\right)} \frac{\Gamma \left(\frac{1}{2} \left(1-i \varpi +\Sq\right)\right)}{\Gamma \left(\frac{1}{2} \left(1-i \varpi -\Sq\right)\right)}  \frac{\Gamma \left(\frac{1}{2} \left(1-\frac{1+\widetilde Q^2}{1-\widetilde Q^2}i \varpi +\Sq\right)\right)}{\Gamma \left(\frac{1}{2} \left(1-\frac{1+\widetilde Q^2}{1-\widetilde Q^2}i \varpi -\Sq\right)\right)} \\
  &\quad- \theta(-\mathrm{Im}\,\varpi)\ \frac{2\pi i\, \xi^\xi  r_h^{-\xi}}{\Gamma\left(\frac{\xi}{2}\right)\Gamma\left(1+\frac{\xi}{2}\right)} \left(\frac{\left(\varpi ^2-q^2\right)}{16}\right)^\frac{\xi}{2}
  \end{split}
\eea
for $0\leq \re \varpi \lesssim \sqrt{1+q^2}$, and
\bea
\begin{split} \label{corrfinal}
 G_\mathrm{reg} &= \frac{2\pi\, \xi^\xi r_h^{-\xi} }{\Gamma\left(\frac{\xi}{2}\right)\Gamma\left(1+\frac{\xi}{2}\right)}\left(\frac{\left(\varpi ^2-q^2\right)}{16}\right)^\frac{\xi}{2} \left[i+\left(\frac{1+i \Sqt}{1-i \Sqt}\right)^\frac{\xi}{2} \frac{e^{-i \xi \Sqt}}{\mathcal{R}}\right]^{-1} \\
 &=  \frac{2\pi\, \xi^\xi r_h^{-\xi} }{\Gamma\left(\frac{\xi}{2}\right)\Gamma\left(1+\frac{\xi}{2}\right)}\left(\frac{\left(\varpi ^2-q^2\right)}{16}\right)^\frac{\xi}{2}  \\
&\quad \times \left[i-\left(\frac{1+i \Sqt}{1-i \Sqt}\right)^\frac{\xi}{2} e^{-i \xi \Sqt}\left(1-\widetilde Q^2\right)^{i\Sqt}\right.\\
& \quad \left.\times \ \frac{\Gamma \left(1-i\Sqt\right)}{\Gamma \left(1+i\Sqt\right)} \frac{\Gamma \left(\frac{1}{2} \left(1-i \varpi +i\Sqt\right)\right)}{\Gamma \left(\frac{1}{2} \left(1-i \varpi -i\Sqt\right)\right)}\frac{\Gamma \left(\frac{1}{2} \left(1-\frac{1+\widetilde Q^2}{1-\widetilde Q^2}i \varpi +i\Sqt\right)\right)}{\Gamma \left(\frac{1}{2} \left(1-\frac{1+\widetilde Q^2}{1-\widetilde Q^2}i \varpi -i\Sqt\right)\right)}\right]^{-1}
\end{split}
\eea
for $\re \varpi \gtrsim \sqrt{1+q^2}$. These results hold up to corrections suppressed by $1/\xi$. The precise regime of validity on the complex $\varpi$ plane of the two expressions is determined by the saddle point approximations discussed in Appendix~C of~\cite{Betzios:2018kwn}. We also adopted a regularization procedure to remove the infinities which occur whenever $\xi$ is an even integer (subtraction of a vacuum AdS correlator), see Appendix~B of the same reference.

\begin{figure}[t]
\centering
\includegraphics[width=0.5\textwidth]{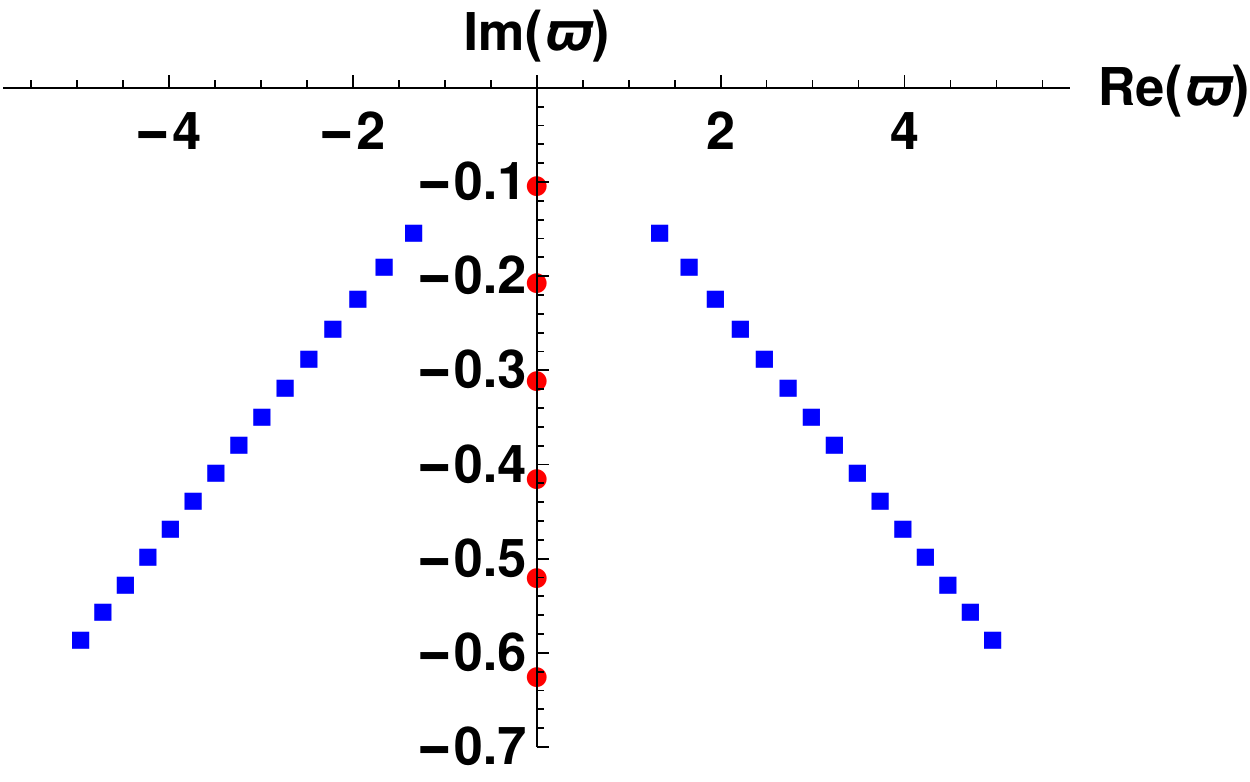}%
\includegraphics[width=0.5\textwidth]{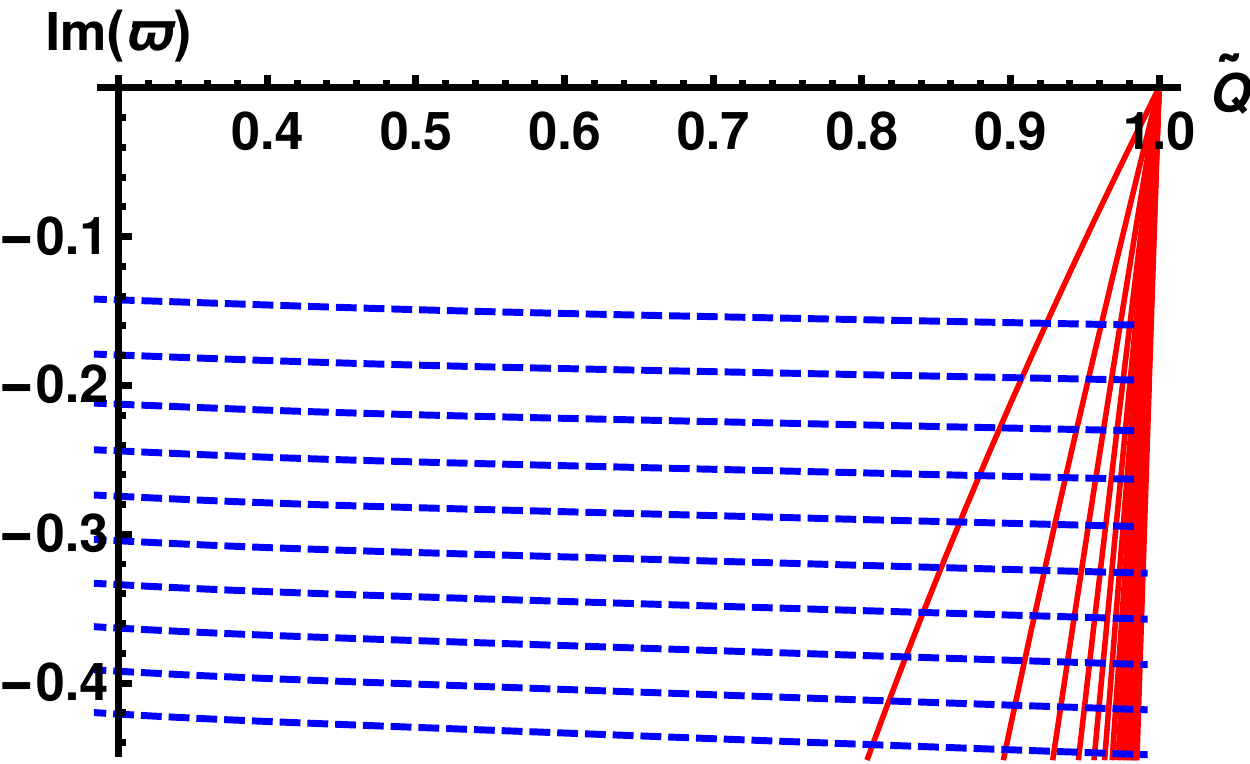}
\caption{
Left: The QNMs of the CR background at $\widetilde Q=0.95$ and $X=-0.47$. Right: The charge dependence of the imaginary parts of the QNMs at $X=-0.47$. The red filled circles/solid curves and blue boxes/dashed curves,  are  the imaginary near-horizon modes (set (iv)) and  the complex CR modes (set (ii)), respectively.
}
\label{fig:immodesCR}
\end{figure}

\begin{figure}[t]
\centering
\includegraphics[width=0.5\textwidth]{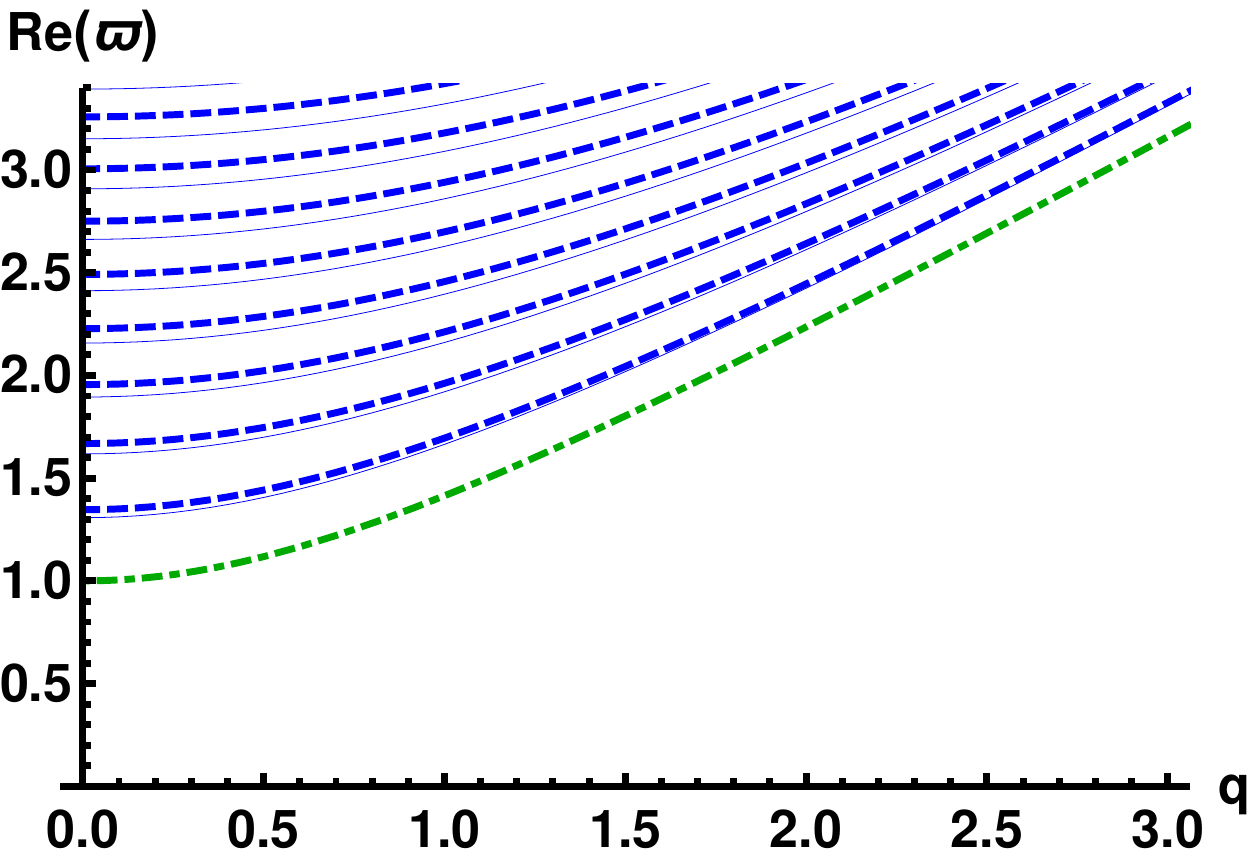}%
\includegraphics[width=0.5\textwidth]{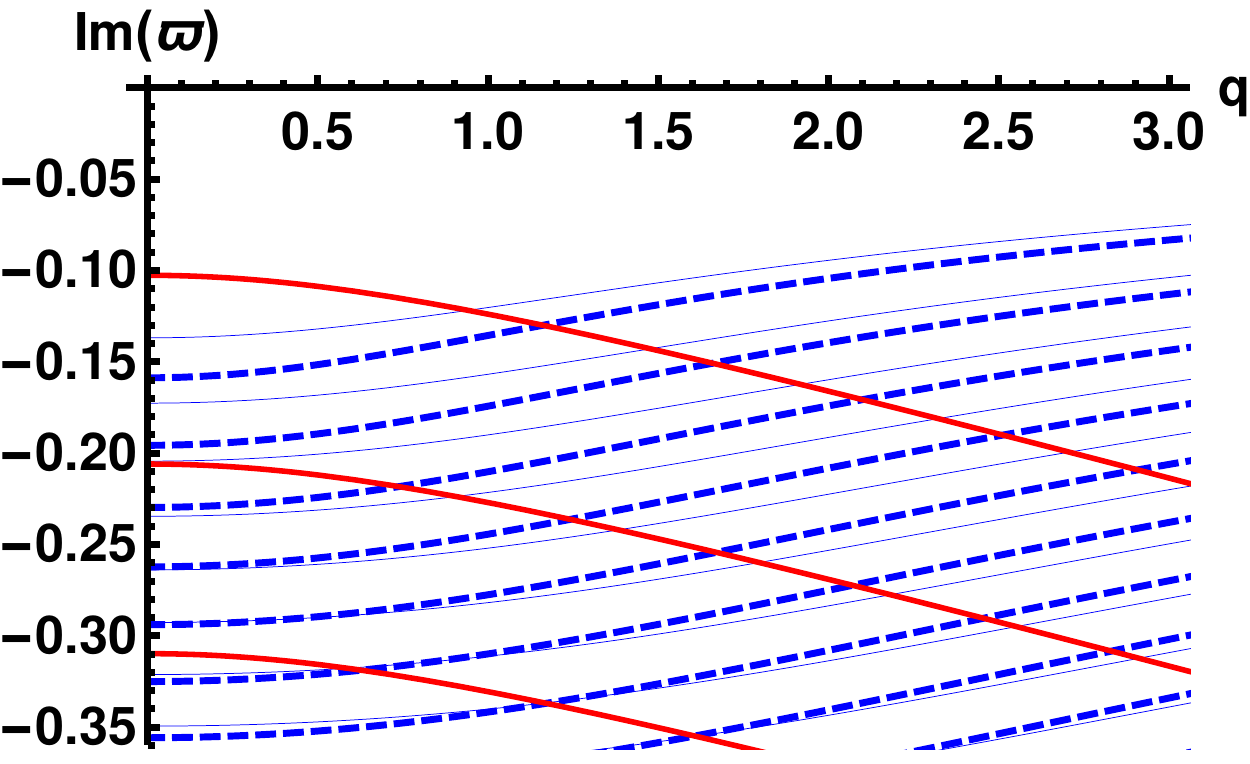}
\caption{
The dependence of the QNMs of the CR background on $q$ at $X=-0.47$. The thick dashed blue and thick solid red curves show our result at $\widetilde Q=0.95$ and the thin solid curves are the result at $\widetilde Q =0$  for comparison. The red curves and blue curves are the imaginary near-horizon modes and  the complex CR modes, respectively. Left: The real parts of the QNMs. The green dotdashed curve is the ``critical'' line $\mathrm{Re}\,\varpi =\sqrt{1+q^2}$. Right: The imaginary parts.
}
\label{fig:qdep}
\end{figure}

The quasi normal modes of the charged CR geometry are obtained as the poles of the analytic expressions~\eqref{corrsmallomega} and~\eqref{corrfinal}. We analyse their behavior in Figs.~\ref{fig:immodesCR} and~\ref{fig:qdep}. There are two sets of modes: the CR modes (i.e., set (ii) defined in the introduction), which arise from the zeroes of the expression in square brackets in the last two rows of~\eqref{corrfinal}, and the imaginary near-horizon modes (set (iv)) which are given directly as the poles of the reflection amplitude as we see directly from~\eqref{corrsmallomega}. Notice that this expression also contains the spurious poles of set (i) which are not true modes of the two-point function.  They appear in~\eqref{corrsmallomega} because the analytic approximation fails in the immediate vicinity of the poles, see~\cite{Betzios:2018kwn} for details. 
{One can actually show that even the solutions to the fluctuation equations remain fully regular near these points, see the discussion in~\cite{Kovtun:2005ev}. These modes have also been analyzed in the literature in the context of little string theory~\cite{Parnachev:2005hh}.
Little string theory actually contains an additional set of singularities, poles of $\Gamma((1-i\varpi-S)/2)$, which were analyzed in detail in~\cite{Bertoldi:2009yi}. However, as we pointed out in~\cite{Betzios:2018kwn}, these latter modes are absent in our results since they are automatically removed by our boundary conditions. Instead of poles, our reflection amplitude has zeroes at these points.  
}

The CR (imaginary) 
modes are shown as blue (red) dots and curves in the plots.  
The imaginary modes can also be solved analytically as we showed above in Eq.~\eqref{immmodesCR}. We emphasize that these are present only at finite charge. 
Notice that the correlator $G_\mathrm{reg}$ depends on $r_h$ only through the proportionality factor. This is a consequence of the scaling property of the fluctuations explained below~\eqref{h2eq}. That is, the correlator and the QNMs only depend on the temperature implicitly through the definitions of $\varpi$, $q$, and $\widetilde Q$.

In Fig.~\ref{fig:immodesCR} (left) we show the locations of the modes at $X=-0.47$ (so that $\xi \approx 26.8$) and at a high value of the charge, $\widetilde Q =0.95$.  Compared to the results at zero charge~\cite{Betzios:2017dol,Betzios:2018kwn}, the locations of CR modes are practically unchanged, so the main difference is the appearance of the imaginary modes.   Fig.~\ref{fig:immodesCR} (right) shows the dependence of the imaginary parts of the rescaled frequencies $\varpi$ on the charge. The imaginary parts of the CR modes (and also the real parts which are not shown) have a mild dependence on charge. For the imaginary modes the dependence of $\mathrm{Im}\,\varpi$ on charge is strong.  As the charge is increased from zero to finite values these modes appear from $\varpi = -i\infty$ and approach the origin in the maximally charged solution. Notice however that the movement of the modes presented here is in part the consequence of our definition of the scaled frequency $\varpi$: we defined it in~\eqref{varpidef} as $\varpi \sim \omega/\xi$ rather than as the ratio $\omega/2\pi T$ (with $T$ given in~\eqref{thermoT}) because this leads to much simpler expressions in our case. If we would have used the more common definition in terms of the temperature, the CR modes would move as the charge varies whereas the frequencies of the imaginary modes would remain roughly constant. 
We also note that 
the dominant mode (having the largest $\mathrm{Im}\varpi$) is a CR mode at low values of the charge but an imaginary mode for near-extremal charged geometries.
This is not surprising but happens typically for near-extremal charged black holes: see, e.g.,~\cite{Cardoso:2017soq}.

The dependence of the modes on $q$ is shown in Fig.~\ref{fig:qdep}. We note that this is affected only a little by charge: the thick lines for near extremal black hole are similar to the neutral case (thin lines). Interestingly, the momentum dependence of $\mathrm{Im}\varpi$ is the opposite for CR and imaginary modes as seen from the right hand plot: $\mathrm{Im}\varpi$ increases (decreases) with increasing $q$ for the CR (imaginary) modes. {In passing, we note that the large $q$ limit of the modes can also be studied using the WKB approximation. This is presented in Appendix C.}

\begin{figure}[t]
\centering
\includegraphics[width=0.7\textwidth]{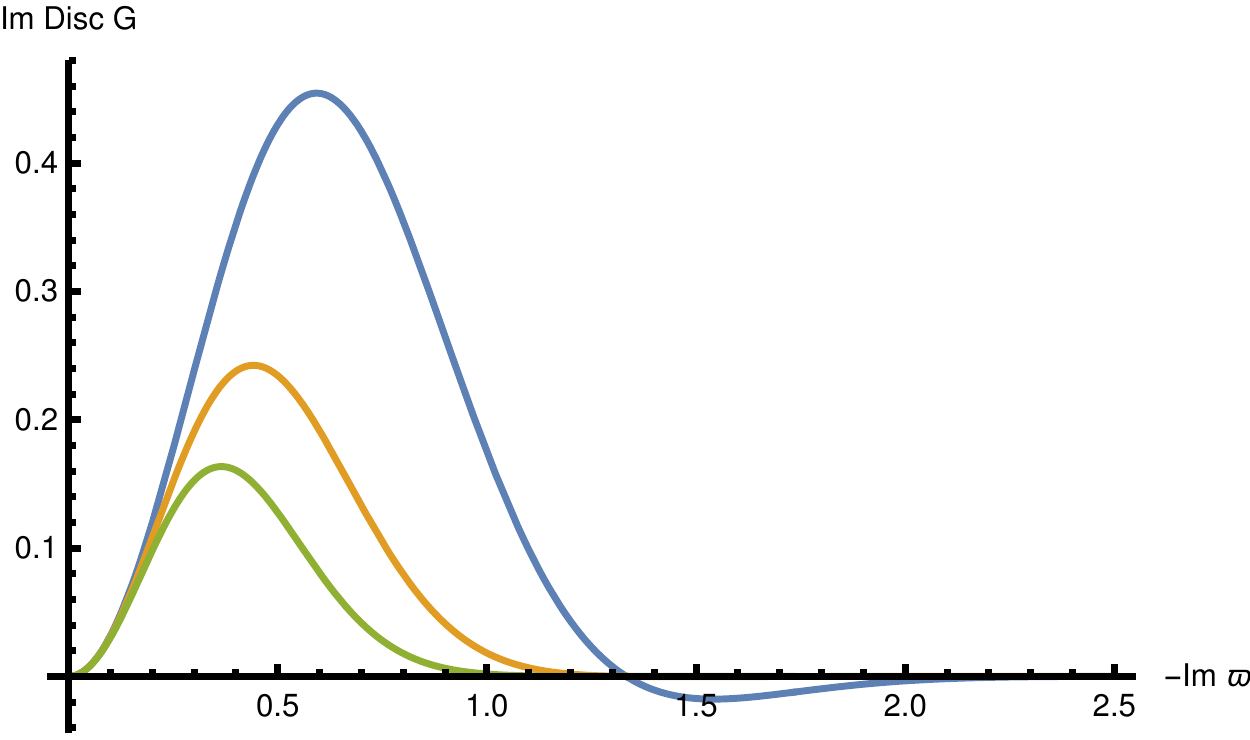} 
\caption{ The discontinuity of the transverse correlator of the energy-momentum tensor on the imaginary axis of the complex frequency plane in the extremal limit $\widetilde Q \to 1$ at $q=0$. The blue, yellow, and green curves are for $\xi = 10$, 20, and 30, respectively.
} 
\label{fig:GDisc}
\end{figure}

As we pointed out above,  the poles on the imaginary axis arising from the reflection amplitude accumulate into a branch cut in the extremal limit (see Eq.~\eqref{imbranchcut}). The resulting discontinuity of the correlator is plotted in Fig.~\ref{fig:GDisc} for $\xi=10$, 20, and 30 and for $q=0$.

\subsubsection{UV completed CR geometry}

In this section we address the pressing issue of how to define the holographic dictionary for the dilatonic black brane backgrounds i.e. the CR background. This is not immediately obvious  as the geometry of the CR background is not asymptotically AdS near the boundary. In~\cite{Gursoy:2015nza} the dictionary was obtained by using generalized  dimensional reduction~\cite{Gouteraux:2011qh} from a higher dimensional AdS background through which the dictionary can be borrowed from the parent AdS solution. A more solid way of ensuring a regular UV behavior, however, is to choose a dilaton potential such that it generates an RG flow from AdS$_5$ in the UV to the CR geometry in the IR. 

In~\cite{Betzios:2017dol,Betzios:2018kwn} the authors introduced an analytic ``approximation'' to such RG flow which captures its main features while maintaining analytic control of the solutions. This was obtained by gluing a slice of the AdS$_5$ geometry directly to the CR geometry. The approach is detailed in Sec.~5 of~\cite{Betzios:2018kwn}. As in the neutral case, one of the main effects of the procedure is that the location of the horizon can no longer be scaled out in the fluctuation equations so that the correlators and QNMs depend nontrivially on the temperature.

Requiring continuity of the fluctuations and their derivatives at the joint between the AdS and CR geometries leads to mixing of the source and vev terms for the two geometries, described in terms of the transition matrix
\begin{align} 
\begin{split}\label{Mglued}
 M_{11} &= \frac{  2 i  \pi \left(\frac{\mu}{2}\right)^{\xi/2} }{\tilde m  \Gamma \left(\frac{\xi }{2}\right)} \left[J_1(\tilde m ) H_{\xi /2}^{(1)}(\mu)-J_2(\tilde m ) H_{\xi/2-1}^{(1)}(\mu)\right] \, ,\\
 M_{12} &= \frac{2  \Gamma \left(\frac{\xi }{2}+1\right)}{ \left(\frac{\mu}{2}\right)^{\xi/2}  \tilde m}\left(\frac{\xi-1}{\xi-4}\right)^\xi \left[J_1(\tilde m) J_{\xi/2}(\mu )-J_2(\tilde m) J_{\xi/2-1}(\mu )\right] \, ,\\ 
 M_{21} &= \frac{ \pi ^2  \left(\frac{\mu}{2}\right)^{\xi/2}  \tilde m^{3} }{16 r_c^4 \Gamma \left(\frac{\xi }{2}\right)}\left[H_1^{(1)}(\tilde m) H_{\xi/2}^{(1)}(\mu )-H_2^{(1)}(\tilde m) H_{\xi/2 -1}^{(1)}(\mu )\right] \, ,\\
 M_{22} &= \frac{i    \pi  \tilde m ^3  \Gamma \left(\frac{\xi }{2}+1\right)}{16 r_c^4\left(\frac{\mu}{2}\right)^{\xi/2} } \left(\frac{\xi-1}{\xi-4}\right)^\xi 
  \left[H_2^{(1)}(\tilde m ) J_{\xi/2-1}(\mu)-H_1^{(1)}(\tilde m ) J_{\xi /2}(\mu)\right] \, ,
  \end{split}
\end{align}
where the subscripts 1 and 2 refer to source and response respectively, 
$r_c$ is the location of the joint, $\tilde m = m r_c$, and $\mu = m r_c (\xi -1)/3$, with $m=\sqrt{\omega^2-k^2}$.

The analytic result for the transverse spin-two correlator for the glued geometry is given in terms of the transition matrix by\footnote{The gluing procedure also implies that the natural choice for the bulk coordinate $r$ changes~\cite{Betzios:2017dol,Betzios:2018kwn}: if we set the UV boundary of the geometry to lie at $r=0$, the coordinate values are shifted by $\Delta r \approx \ell'$ in the CR part of the geometry due to the gluing.} 
\be \label{Gtildetext}
 \widetilde G_\mathrm{reg}  = \frac{M_{21}+M_{22}  G_\mathrm{reg}}{M_{11}+M_{12}  G_\mathrm{reg}} \ ,
\ee
which is regulated in the same way as $G_\mathrm{reg}$ above, given in Eqs.~\eqref{corrsmallomega} and~\eqref{corrfinal}, see Appendix~E in~\cite{Betzios:2018kwn}. 
Notice that $G_\mathrm{reg}$ in~\eqref{corrsmallomega} and in~\eqref{corrfinal} depends on the frequency and momenta only through the rescaled quantities $\varpi$ and $q$. The location of the horizon $r_h$ only appears explicitly in the overall factor, so that the dependence on the temperature is trivial. This is no longer true for $\widetilde G_\mathrm{reg}$ in~\eqref{Gtildetext} both because the matrix elements $M_{ij}$ depend on the location of the horizon and because the overall factor of $G_\mathrm{reg}$ can no longer be factored out. This gives rise to a nontrivial dependence of the QNMs on the temperature. At small temperatures (and not too high frequencies) the QNMs however agree with those obtained in the absence of the UV completion, and are therefore given directly by $G_\mathrm{reg}$, as we have shown in~\cite{Betzios:2018kwn}.

This UV completion with direct gluing of the CR and AdS geometries in the UV neglects nontrivial effects from the blackening factor. We analyzed these effects in~\cite{Betzios:2018kwn} and showed that they are small provided that $\xi (T_c-T)/T_c \gg 1$, where 
\be \label{eq:Tcdef}
 T_c = \frac{3}{4\pi r_c}\left(1-\widetilde Q^2\right) \ .
\ee
That is, the result is valid for $\xi \gg 1$ and if temperature is smaller than $T_c$ by a margin which is much larger than $1/\xi$. As explained in~\cite{Betzios:2017dol,Betzios:2018kwn}, the critical temperature $T_c$ can be understood to be the temperature of the linear dilaton background\footnote{It is not obvious that the expressions for the temperature of the linear dilaton background in~\eqref{eq:Tcdef} and~\eqref{thermoTs} are equivalent. However for this purpose it is enough that they match for $\xi \to \infty$ and $T \approx T_c$, in which case~\cite{Betzios:2018kwn} we find that $r_c \approx \ell$ and $r_h \approx \ell'$ (we have set the coefficient $A_0$ of~\cite{Betzios:2018kwn} to zero).}.

\begin{figure}[t]
\centering
\includegraphics[width=0.5\textwidth]{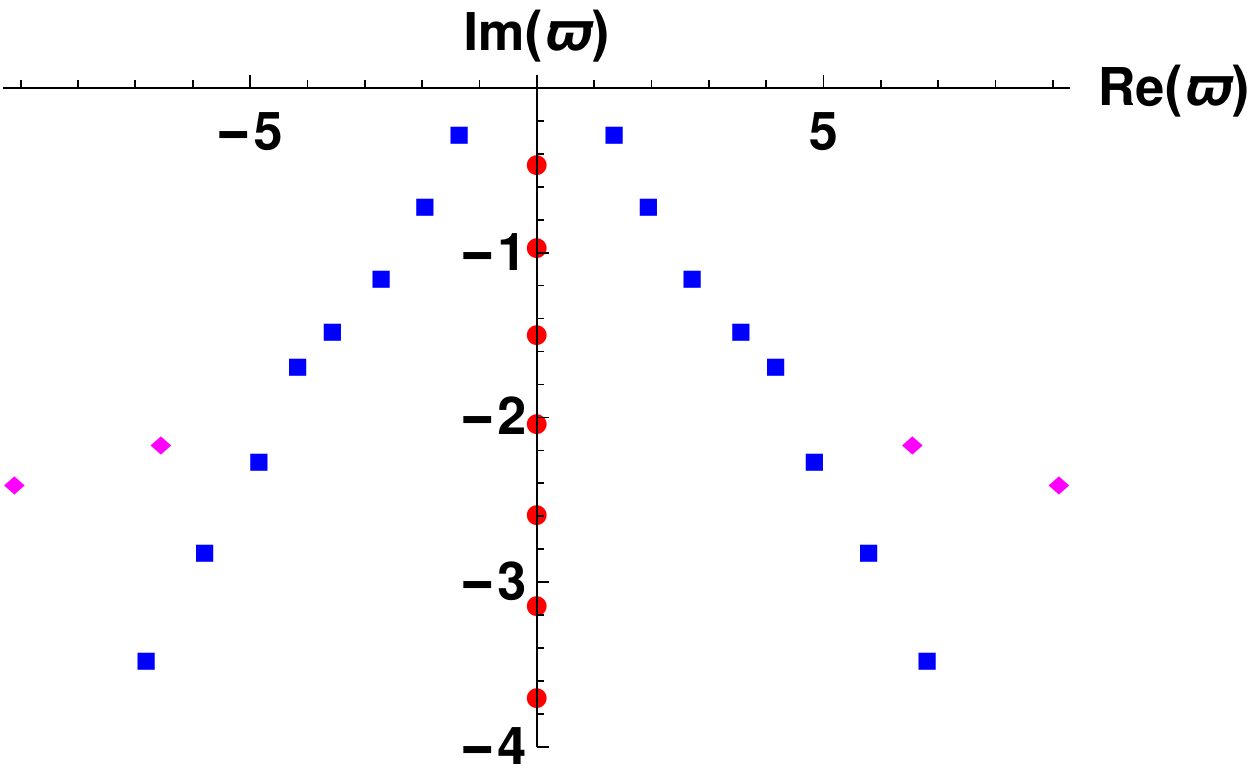}%
\includegraphics[width=0.5\textwidth]{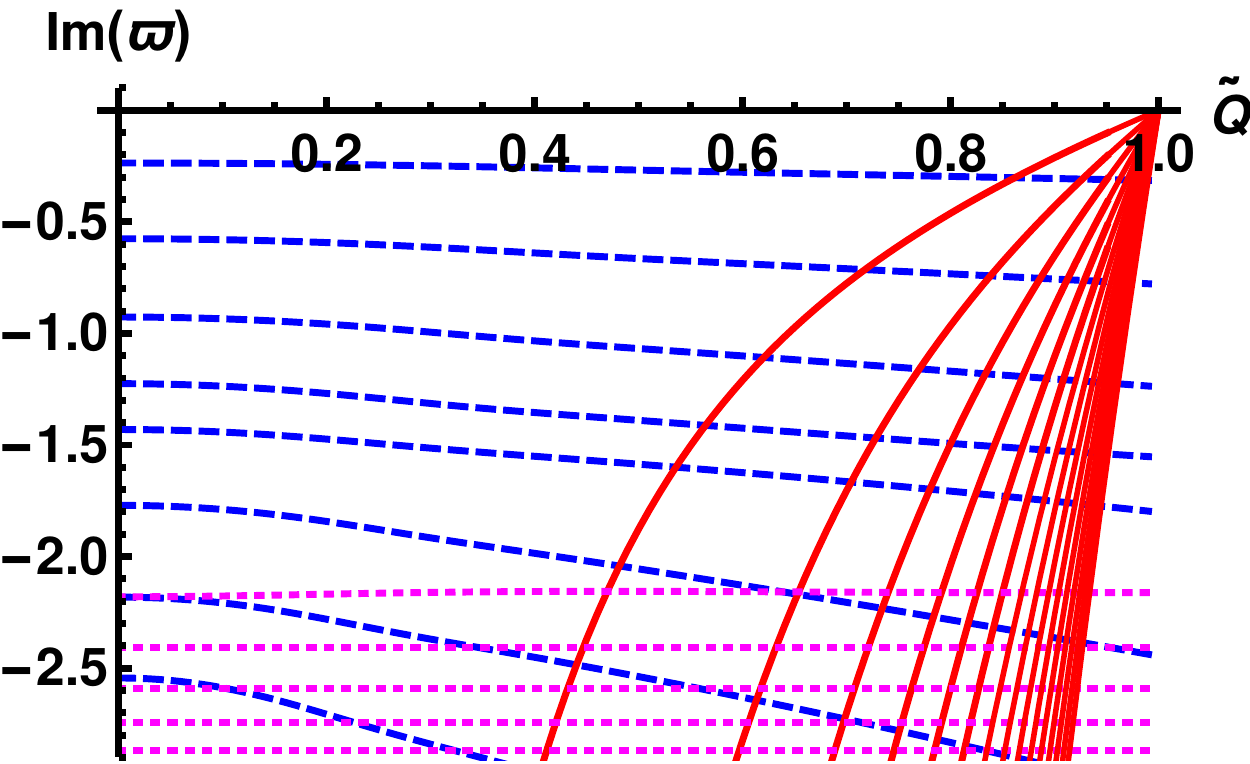}
\includegraphics[width=0.5\textwidth]{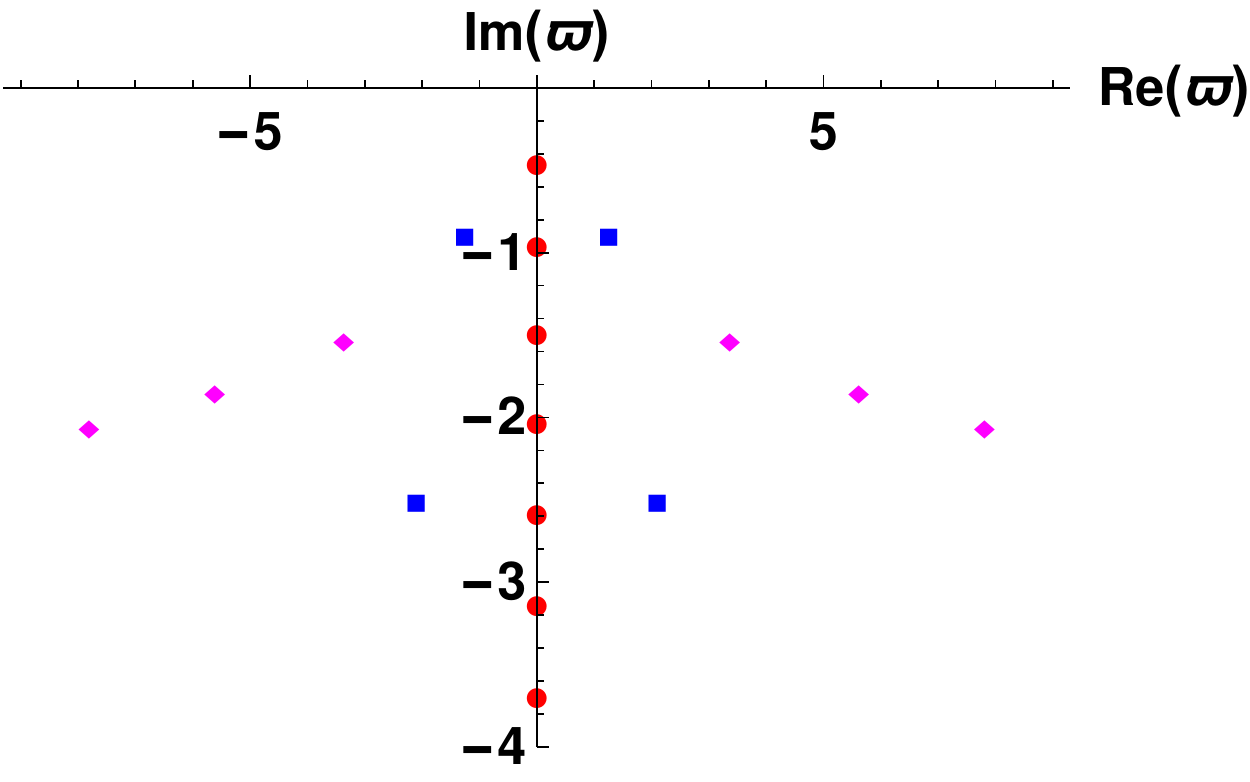}%
\includegraphics[width=0.5\textwidth]{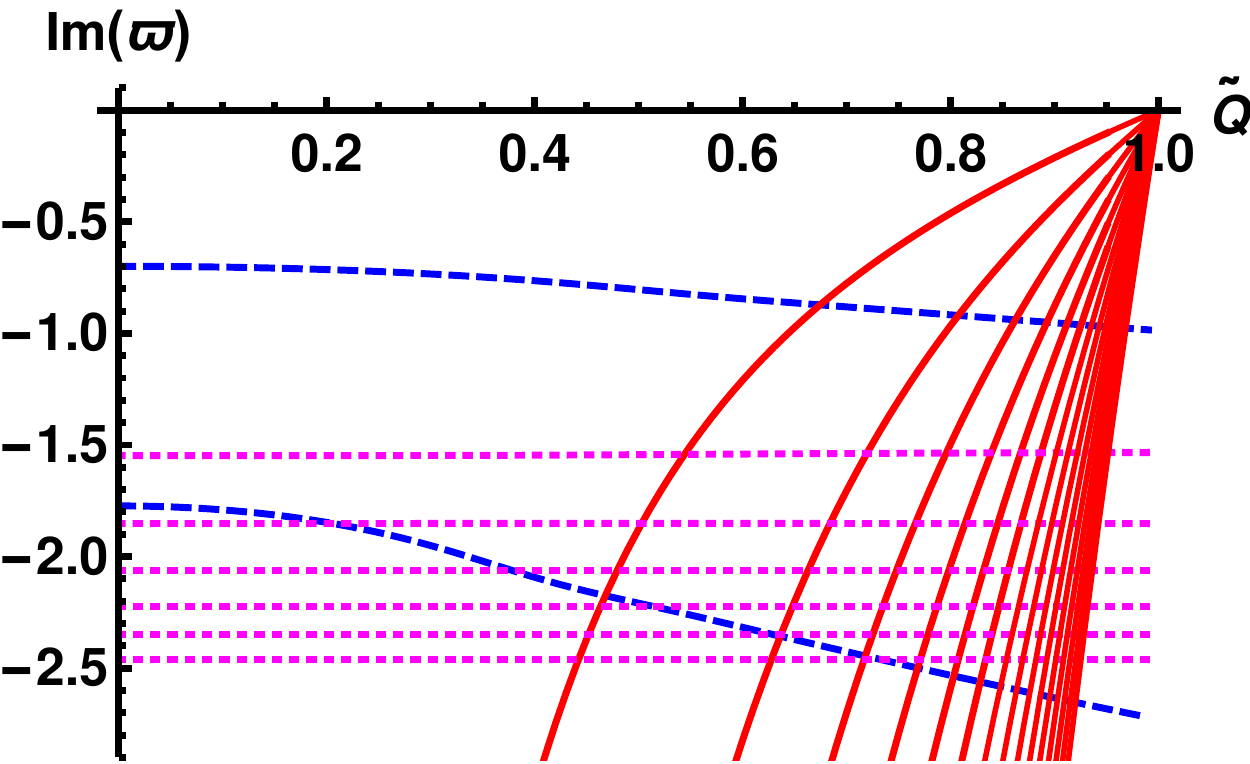}
\caption{
Left column: The QNMs of the glued background at $\widetilde Q=0.8$ and $X=-0.47$. Right column: The charge dependence of the imaginary parts of the QNMs at $X=-0.47$. Top row: intermediate temperature, $T_c/T =1.4$. Bottom row: high temperature, $T_c/T =1.2$. 
The red filled circles/solid curves, blue boxes/dashed curves, and magenta diamonds/dotted curves are the imaginary near-horizon modes (set(iv)), the CR modes (set (ii)), and the AdS modes (set (iii)), respectively.
}
\label{fig:immodes}
\end{figure}

\begin{figure}[t]
\centering
\includegraphics[width=0.5\textwidth]{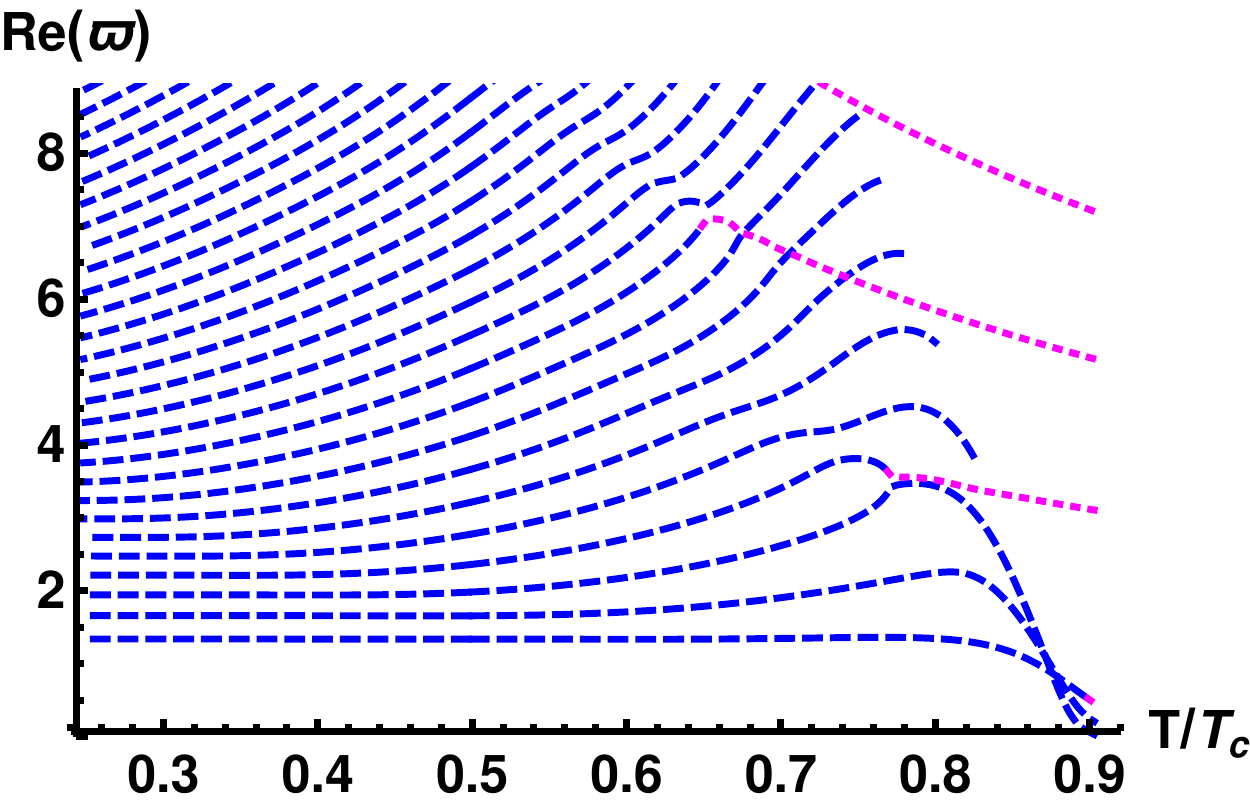}%
\includegraphics[width=0.5\textwidth]{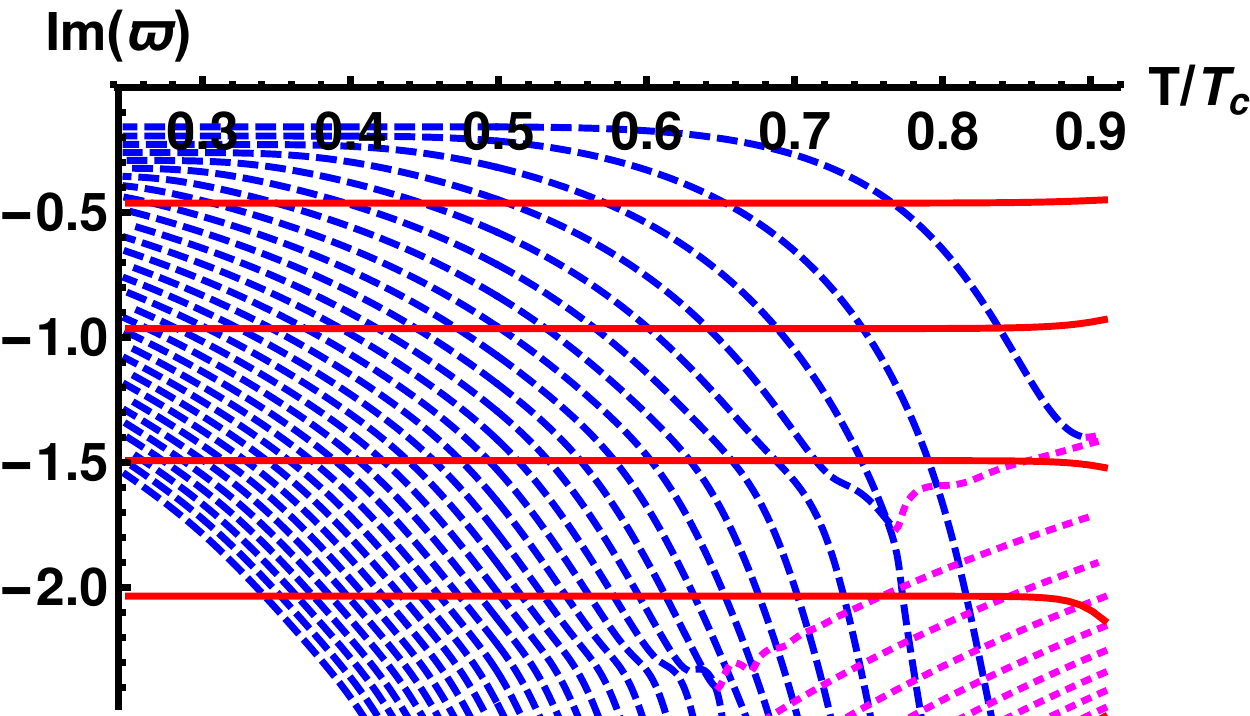}
\caption{
The temperature dependence of the QNMs in the glued background at $\widetilde Q=0.8$ and $X=-0.47$. Left: real parts. Right: imaginary parts. The red solid curves, blue dashed curves, and the magenta dotted curves are the imaginary near-horizon modes, the complex modes, and the AdS modes, respectively.
}
\label{fig:Tdep}
\end{figure}

The quasi normal modes can be obtained as the poles of the correlator~\eqref{Gtildetext}. We can identify three distinct class of modes (see Fig.~\ref{fig:immodes}):
\begin{itemize}
 \item The AdS modes (magenta dots and curves in Figs.~\ref{fig:immodes} and~\ref{fig:Tdep}) which are controlled by the details of the geometry near UV and only appear after the UV completion. This is the set (iii) of modes as defined in the introduction.
 \item The CR modes (blue dots and curves) which are controlled by the CR geometry but relatively insensitive to the details in the immediate vicinity of the horizon.
  This is the set (ii).
 \item The imaginary near-horizon modes (red dots and curves) which are only present at finite charge. This is the set (iv).
\end{itemize}

In Fig.~\ref{fig:immodes} we show the locations of the QNMs on the complex frequency plane for $\widetilde Q = 0.8$ and at intermediate temperature ($T_c/T = 1.4$, top left plot) and at high temperature ($T_c/T =1.2$, bottom left plot). Both sets of complex modes, i.e., the CR modes and AdS modes, lie roughly on straight lines. The dependence of the imaginary part of the frequencies on charge is shown on the right column in Fig.~\ref{fig:immodes}, for intermediate (top plot) and high temperatures (bottom plot). We observe the AdS UV modes (dashed magenta curves) are essentially independent of charge, whereas the CR modes have mild charge dependence, and the imaginary near-horizon modes have a strong dependence on the charge\footnote{We emphasize that these conclusions are based on our definition of $\varpi$ above.}. This result is expected as in our conventions the charge only affect the near horizon geometry, and in particular the UV completion is independent of charge. The dominant mode, i.e. the mode with largest $\mathrm{Im}\varpi$, is either one of the complex CR modes (low values of the charge) or one of the imaginary near-horizon modes (large values of the charge). The value of the charge where the crossing between the highest CR and imaginary mode takes place also depends on the temperature. Notice that we only included the first few modes in each sector, some of the higher AdS and near-horizon modes would also appear on the right hand plots.

Our results for the CR and near-horizon modes are similar to the ones in pure CR geometry, see Fig.~\ref{fig:immodesCR}. It appears that the UV completion (i.e., temperature effects) leads to slightly stronger charge dependence than for the pure CR case, but the CR modes are also shifted downwards on the complex frequency plane with respect to the results for the pure CR geometry. We also found a mild dependence of the real parts of the CR and AdS modes on charge (not shown in these plots). 
One can check that in the limit of zero temperature the plots approach those of Fig.~\ref{fig:immodesCR}. 

Some effects due to finite temperature are already visible in Fig.~\ref{fig:immodes}: the CR modes move downwards with increasing temperature, the AdS modes move upwards, whereas the imaginary near-horizon modes stay unaffected. We study the temperature dependence in more detail in Fig.~\ref{fig:Tdep}. From the left hand plot we also see that the CR (AdS) modes move to larger (smaller) $|\mathrm{Re}\varpi|$ as $T$ grows. Furthermore, we observe that the CR and AdS modes interact nontrivially as the temperature varies. This can also be seen by comparing the plots in the left column of Fig.~\ref{fig:immodes}: as the line of the CR mode moves down, additional AdS modes appear. Some of the blue dashed curves of the CR modes end without exiting the plot at rather high temperatures in Fig.~\ref{fig:Tdep} (left) because the imaginary parts of the modes grow so large that tracking them is difficult.

While the complex modes exhibit moderate dependence on temperature, the imaginary near-horizon modes are almost temperature independent. This is expected since the nontrivial temperature effects only arise in our setup from the UV completion.


Finally, we note that, while the correlator $G_\mathrm{reg}$ in~\eqref{corrsmallomega} and in~\eqref{corrfinal} is not smooth in the limit $\xi \to \infty$ 
the UV completion through gluing the geometry with a slice of AdS$_5$ near the boundary removes this problem. Following the steps in Sec.~5.4 of~\cite{Betzios:2018kwn}, we obtain in the limit $\xi \to \infty$ the following transition matrix:
\begin{align}
\begin{split} \label{MgluedX12}
 M_{11} &= \frac{4 e^{\frac{3}{2} \left(1-\sqrt{1-\hat \mu ^2}\right)} \left[\hat \mu  J_1\left(\frac{3 \hat \mu }{2}\right)-\left(1-\sqrt{1-\hat \mu ^2}\right) J_2\left(\frac{3 \hat \mu }{2}\right)\right]}{3 \hat \mu ^2} \, ,\\
 M_{12} &= \frac{4 e^{\frac{3}{2} \left(1+\sqrt{1-\hat \mu ^2}\right)} \left[\hat \mu  J_1\left(\frac{3 \hat \mu }{2}\right)-\left(1+\sqrt{1-\hat \mu ^2}\right) J_2\left(\frac{3 \hat \mu }{2}\right)\right]}{3 \hat \mu ^2}
 \,,\\
 M_{21} &= -\frac{27 i \pi  e^{\frac{3}{2} \left(1-\sqrt{1-\hat \mu ^2}\right)} \hat \mu ^2 \left[\hat \mu  H_1^{(1)}\left(\frac{3 \hat \mu }{2}\right)-\left(1-\sqrt{1-\hat \mu ^2}\right) H_2^{(1)}\left(\frac{3 \hat \mu }{2}\right)\right]}{128 r_c^4} \,,\\
 M_{22} &= -\frac{27 i \pi  e^{\frac{3}{2} \left(1+\sqrt{1-\hat \mu ^2}\right)} \hat \mu ^2 \left[\hat \mu  H_1^{(1)}\left(\frac{3 \hat \mu }{2}\right)- \left(1+\sqrt{1-\hat \mu ^2}\right) H_2^{(1)}\left(\frac{3 \hat \mu }{2}\right)\right]}{128 r_c^4}  \, ,
 \end{split}
\end{align}
where $\hat \mu = 2 m r_c/3$ which equals $\sqrt{\varpi^2-q^2}$ for geometries with a horizon. Using this matrix, the regulated correlator~\eqref{Gtildetext} at $\xi=\infty$ simplifies to
\begin{align} 
\begin{split}\label{xiinftyG}
&\widetilde G_\mathrm{reg} = -\frac{81 i \pi  \hat \mu^4}{512 r_c^4 } \\
&\times \frac{\hat \mu  \left[1+e^{3 \Sq\le(1-\frac{r_h}{r_c}\ri)} \mathcal{R}\right]H_1^{(1)}\left(\frac{3 \hat \mu}{2}\right)+ \left[(\Sq-1) -e^{3 \Sq\le(1-\frac{r_h}{r_c}\ri)} (\Sq+1) \mathcal{R}\right]H_2^{(1)}\left(\frac{3 \hat \mu}{2}\right)}{\hat \mu  \left[1+e^{3 \Sq\le(1-\frac{r_h}{r_c}\ri)} \mathcal{R}\right]J_1\left(\frac{3 \hat \mu}{2}\right)+ \left[(\Sq-1)-e^{3 \Sq\le(1-\frac{r_h}{r_c}\ri)} (\Sq+1) \mathcal{R}\right]J_2\left(\frac{3 \hat \mu}{2}\right)} \ . 
\end{split}
\end{align}
Unlike the expressions~\eqref{corrsmallomega} and~\eqref{corrfinal} this expression is valid in the whole complex frequency plane. Despite the presence of square roots there are no branch cuts in this expression as the cuts cancel in the full expression. Notice also that while the formula has the same form as in the neutral case of~\cite{Betzios:2018kwn}, but charge dependence arises through the modified definitions of $\varpi$, $q$, and the reflection amplitude $\mathcal{R}$ in~\eqref{Refdef}.

In the $\xi \to \infty$ limit we expect that the CR modes accumulate to form a branch cut on the real axis of the frequency plane~\cite{Betzios:2017dol,Betzios:2018kwn}. Due to the UV completion the result~\eqref{xiinftyG} however still has discrete QNMs rather than a branch cut, but the result has smooth limit $r_h \to \infty$ which suppresses the effect of the UV completion on the QNMs and the branch cut reappears. We obtain the following result for the discontinuity at the branch cut, when $|\hat \mu|>1$,
\be \label{discxiinfty}
 \mathrm{Disc}\lim_{r_h\to\infty}\widetilde G_\mathrm{reg}= \frac{27 i \hat \mu ^3 \sqrt{\hat \mu ^2-1}}{64 \left(\hat \mu  \left(J_1\left(\frac{3 \hat \mu }{2}\right)\right)^2-2 J_2\left(\frac{3 \hat \mu }{2}\right) J_1\left(\frac{3 \hat \mu }{2}\right)+\hat \mu  \left(J_2\left(\frac{3 \hat \mu }{2}\right)\right)^2\right)} \ ,
\ee 
where $\hat\mu = \sqrt{\varpi^2-q^2}$. This limiting expression again has the same for as at zero charge, no longer contains even the reflection amplitude, so that the dependence on charge only appears through the modified definitions of $\varpi$ and $q$ in~\eqref{varpidef}. Recall that these branch cuts, which appear on the real axis for $\xi \to \infty$, are therefore of different nature than the branch cut of Fig.~\ref{fig:GDisc}, which appears on the imaginary axis in the extremal limit. 
Actually, as the result~\eqref{discxiinfty} is independent of the reflection amplitude, it does not carry any information of the imaginary modes appearing at finite charge. These modes are still present in~\eqref{xiinftyG}, but their residues are exponentially suppressed as $r \to r_h$. That is, these modes are decoupled in the $\xi \to \infty$ limit, but present at finite, large $\xi$. The same can be observed by studying numerically the full expression~\eqref{Gtildetext}: the residues of the imaginary modes are suppressed as $\xi \to \infty$, unless one takes $T \to T_c$ at the same time.




\subsection{Analytic fluctuations: WKB approach} \label{sec:WKB}


In this subsection we discuss another complementary approach for computing the transverse correlators of the energy-momentum tensor analytically, which is in part based on the WKB approximation and we apply this alternative method to compute the quasi-normal spectrum of the so-called hyperscaling violating Lifshitz BH solutions. This alternative approach is applicable to a general class of geometries for the following reasons. In the approach discussed above and in~\cite{Betzios:2017dol,Betzios:2018kwn} we did not make use of the fact that $\xi$ is large in the near boundary solutions even though our final results are valid only to leading order in $1/\xi$. We did this because for the geometries we discussed, which match with zero temperature CR geometries in the near boundary regime,  fluctuation equations could be solved analytically in terms of Bessel functions for all values of $\xi$ in this regime (see~\eqref{bdrysol}) and there was no need to consider the limit of large $\xi$. For more general geometries such analytic solutions will not be known, but the fluctuation equations can still be solved by employing large $\xi$ approximation: this leads to the WKB approach which we discuss in this section.  We stress that this only changes the treatment near the boundary, and the solutions in the IR near the horizon need to be analyzed as above. We recall that the charge dependence above arose solely from the IR regime.

We will apply this WKB approach to a different class of solutions  to the Einstein-Maxwell-Dilaton action than the solution discussed in Sec.~\ref{sec:chargedCR}, namely the hyperscaling violating Lifshitz solutions. These solutions have been discussed extensively in the literature~\cite{Gouteraux:2011ce,Huijse:2011ef,Gath:2012pg} 
mainly in connection to applications of holography in condensed matter systems, and indeed belong to the class of geometries where the UV fluctuations cannot be (except for specific cases) solved analytically, so that the method of~\cite{Betzios:2018kwn} cannot be used to analyze the correlators directly. These geometries solve the EMD system for a choice of different set of exponents in the potentials: 
\be
 V(\phi) = V_0 e^{-\frac{8 X}{3}\phi} \ , \qquad Z(\phi) = e^{\frac{8X'}{3}\phi} \ ,
\ee
where we do not set $X=X'$ unlike above. In the Eddington-Finkelstein coordinates a special hyperscaling violating Lifshitz solution reads
\bea 
\begin{split}\label{metric2}
 ds^2 &= \le(\frac{ r}{ \ell}\ri)^{ \delta+\frac{2}{3} (-\hat\xi +1)}\left[-2 d r dv - \left(1-\left(\frac{ r}{ r_h}\right)^{ \hat\xi }\right) dv^2\right] +  \le(\frac{ r}{ \ell}\ri)^{\frac{2}{3} (-\hat\xi +1)}\delta_{ij} dx^{i} dx^{j} \, ,\\
 \phi(r) &= \frac{1}{3} \le(X-X'\ri)(3 \delta- \hat\xi +4) \log \left(\frac{r}{\ell}\right) \,,\\
 A_t(r) &=\sqrt{-\frac{ \delta}{ \hat\xi}} \left(\frac{r}{\ell}\right)^{- \hat\xi } \left(1-\left(\frac{ r}{ r_h}\right)^{ \hat\xi }\right)\,, 
\end{split}
\eea
where we defined the constants 
\bea
\begin{split}
  \hat\xi &=
  \frac{8 (X-X') (X+2 X')-9}{4 (X-X') (5 X+X')-9}\, ,\\
  \delta &= 
  \frac{6 (4 X (X'-X)+3)}{4 (X-X') (5 X+X')-9} \ ,
  \end{split}
\eea
and $\ell$ is defined as
\be
 \ell^2 V_0 =   \hat\xi  \le( \hat\xi -1-\frac{ \delta}{2}\ri) \ .
\ee
Notice that the parameter $\hat\xi$ plays a similar role as the parameter $\xi$ defined in~\eqref{xidef} for the charged CR geometries. In particular, we will be able to derive the QNMs analytically in the limit $\hat\xi \to \infty$. Its definition however differs from that of $\xi$ even in the limit $X' \to X$ which was assumed in the charged CR case. We will drop the hat for notational simplicity below.

These are ``scaling'' geometries, i.e., the scale transformations $r \to \Lambda r$ act on the solution homogeneously as a power law hence they can be viewed as realizing a generalized conformal symmetry in the dual QFT. In particular, under the transformation $x_i \to \lambda x_i$, $v \to \lambda^{1/(1+\delta/2)} v$, $r \to \lambda^{1/(1+\delta/2)} r$ the metric transforms as $ds^2 \to \lambda^{2+\frac{2}{3}\frac{1-\xi}{1+\delta/2}} ds^2$.
The entropy and the temperature are given by
\be
 s= \frac{1}{4 G_5} \le(\frac{r_h}{\ell}\ri)^{1-\xi} = \frac{1}{4 G_5} \le(\frac{4\pi\ell T}{\xi}\ri)^{\xi-1}\ , \qquad\qquad  T =  \frac{ \xi  }{4 \pi  r_h}\ .
\ee
Comparing to the standard formulas for the scaling exponents used in the literature i.e. $v \to \lambda^z v$, $ds^2 \to \lambda^{2\theta/d} ds^2$ (with $d=3$), we  find the following relations 
\be
 z = \frac{1}{1+\delta/2} \ ,\qquad \theta = 3 + \frac{1-\xi}{1+\delta/2}
\ee
so that
\be\lab{parredef}
 \delta = -\frac{2(z-1)}{z}\ , \qquad  \xi -1 = \frac{3-\theta}{z} \ . 
 \ee 
This geometry differs from that of~\eqref{metricCR}--\eqref{gaugeCR} at zero charge ($q_0=0$) by the simple additional warp factor $\sim r^{\delta}$. In particular the near horizon behavior, which is determined by $\xi$ or $\hat\xi$, is similar. One might even expect that the charged counterpart of the geometry~\eqref{metric2} is obtained by modifying the blackening factor as in~\eqref{metricCR}--\eqref{gaugeCR}.
%
This however is not 
the case -- the Lifshitz solution for any charge apparently cannot be solved analytically.\footnote{Nontrivially charged analytic Lifshitz solution can be found for a system with two gauge fields having different potentials $Z(\phi)$~\cite{Tarrio:2011de}.} 
This is why we restrict here to the neutral case.


\subsubsection{Fluctuations in the helicity two sector}

The fluctuation equation for the helicity two modes  using the metric as in~\eqref{metric2}  (which also equals the massless scalar equation in this metric), is given in the Eddington-Finkelstein coordinates as 
\be
 r f(r) \phi ''(r)- \left(\xi -2 i r \omega -f(r)\right)\phi '(r)- \left(k^2 r \left(\frac{r}{\ell}\right)^{\delta }+i (\xi -1) \omega \right)\phi (r)   = 0 
\ee
where $f(r) = 1 -(r/r_h)^\xi$. The parameters $\delta$ and $\xi$ are related to the Lifshitz parameters $z$ and $\theta$ through eqs. (\ref{parredef}). 

 We note that at $k=0$ the dependence on $\delta$ disappears. 
As above, the strategy to calculate the two-point function analytically at large $\xi$ is to solve the equation in the regimes (near the UV boundary and near the horizon) analytically and match the solutions in the middle.
As it turns out, it is enough to consider real $\varpi$: the results at real values can be continued to the whole complex plane in the same way as in Sec.~\ref{sec:ancorrs}.

\subsubsection{Solutions near the boundary at large $\xi$}

In order to study the solutions of the fluctuation equation near the boundary, it is useful to study the limit $\xi \to \infty$ at fixed $r$. More precisely, we define
\be
 \varpi = \frac{\omega}{2\pi T} = \frac{2 \omega r_h} {\xi} \ , \qquad q = \frac{k}{2\pi T} = \frac{2 k r_h}{\xi}
\ee
and write the fluctuation equation in a Schr\"odinger form by defining $h(r)=r^{-\xi/2}e^{i\omega r}\phi(r)$ and switching to the variable 
\be
y=-\log (r/r_h)\ . 
\ee
This leads to
\be \label{schrode}
 h''(y) - \frac{\xi^2}{4}\le[1 - \varpi^2e^{-2y} + \le(\frac{r_h}{\ell}\ri)^\delta q^2 e^{-(2+\delta)y}  \ri] h(y) = 0 \ ,
\ee
where we also dropped the terms $\propto (r/r_h)^\xi$ which are only important near the horizon.
At large $\xi$ this is solved by the WKB approximation:
\be \label{WKB}
 h(y) =  \frac{C_+}{v(y)^{1/4}} \exp\le(\int_0^y d\tilde y \frac{\xi}{2} \sqrt{v(\tilde y)} \ri) +  \frac{C_-}{v(y)^{1/4}} \exp\le(-\int_0^y d\tilde y \frac{\xi}{2} \sqrt{v(\tilde y)} \ri)\, ,
\ee
where
\be
 v(y) = 1 - \varpi^2e^{-2y} + \le(\frac{r_h}{\ell}\ri)^\delta q^2 e^{-(2+\delta)y} \ ,
\ee
and corrections are suppressed by $1/\xi$.

The computation then proceeds in different ways depending on whether $v(y)$ has a root at positive $y$ or not. We will discuss first the case with root, which means that $|\varpi|>\sqrt{1+q^2(r_h/\ell)^{\delta}}$, corresponding to the case of~\eqref{corrfinal} with the more complicated structure of the correlator.
The essential task is to combine the solutions near the boundary, $y \to \infty$, where $v(y)>0$, to solutions near the horizon $y \to 0$, where $v(y)<0$, in the interesting regime of frequencies where QNMs are found. This cannot yet be done using the WKB result because it is not analytic\footnote{One may wonder whether the branch point can be avoided by going to complex $\varpi$. We will not address this here and take a different route.} when $v(y)=0$. That is, the WKB approximation breaks down near the point $y=y_0$ where $v(y)$ vanishes. In order to see how the different branches of solutions are connected, we expand around $y=y_0$:
\be
v(y)\simeq  v'(y_0)(y-y_0) \ ,
\ee
where $v'(y_0)>0$. Then the fluctuations are solved in terms of the Airy functions:
\be \label{airy}
 h(y) = C_A \mathrm{Ai}\le(\frac{(y-y_0)\sqrt[3]{v'(y_0)\xi^2}}{2^{2/3}}\ri) + C_B \mathrm{Bi}\le(\frac{(y-y_0)\sqrt[3]{v'(y_0)\xi^2}}{2^{2/3}}\ri) \ .
\ee
Matching this with the WKB result we can solve how the various branches of WKB solutions are connected. The matching is carried out in Appendix~\ref{app:WKB}.

\subsubsection{Solutions near the horizon}


In order to solve the fluctuations near the horizon, we define the variable
\be \label{wdef2}
 w = \le(\frac{r}{r_h}\ri)^\xi \ .
\ee
Notice that keeping this variable fixed at large $\xi$ forces $r \to r_h$ so that we will indeed zoom in the near horizon region.
Dropping subleading corrections at large $\xi$ we find for the fluctuation equation
\be
w^2\left(1-w\right) \phi ''(w)+\left(-w^2+i w \varpi \right) \phi '(w) + \left(-\frac{1}{4} q^2 \left(\frac{r_h}{\ell}\right)^{\delta }-\frac{i \varpi }{2}\right)\phi (w) =0 \ .
\ee
This can be solved in terms of hypergeometric functions:
\bea
\begin{split}\label{horsol}
\phi(w) &=C_+^\mathrm{IR}\, w^{\frac{1}{2} \left(1-i \widetilde S-i \varpi\right)}\\
 &\times\, _2F_1\left(\frac{1}{2} \left(-i \varpi -i \widetilde S+1\right),\frac{1}{2} \left(-i \varpi -i \widetilde S+1\right);1-i \widetilde S;w\right) \\
&+C_-^\mathrm{IR}\, w^{\frac{1}{2} \left(1+i \widetilde S-i \varpi\right)}\\
 & \times\, _2F_1\left(\frac{1}{2} \left(-i \varpi +i \widetilde S+1\right),\frac{1}{2} \left(-i \varpi +i \widetilde S+1\right);1+i \widetilde S;w\right)
 \end{split}
\eea
where 
\be \label{Sdef}
 \widetilde S = \sqrt{\varpi^2-\tilde q^2-1} \ , \qquad \tilde q^2 = \le(\frac{r_h}{\ell}\ri)^\delta q^2 \, .
\ee
Matching these with the WKB result~\eqref{WKB} and with~\eqref{airy} gives us full analytic control of the fluctuations at large $\xi$. This matching is carried out in detail in Appendix~\ref{app:WKB}.



\subsubsection{Correlator} 

The final result for the correlator  of the transverse components of the energy-momentum tensor in the WKB approach (to leading nontrivial order in $1/\xi$ and in the large-$\varpi$ region) can be read off from the matched solution constructed in Appendix~\ref{app:WKB}. Combining the results in~\eqref{reflampl}, \eqref{firstmatching}, \eqref{secondtmatching}, \eqref{thirdmatching}, and \eqref{UVexpansion} yields the following analytic expression for the correlator:
\be \label{Glargeom}
 G =  r_h^{-\xi } \exp\Big[\xi \le( y_0- I_2(\varpi,\tilde q)\right)\Big] \left\{\frac{i}{2}+\frac{1}{i+\exp\le[-i \xi I_1(\varpi,\tilde q)\ri] \mathcal{R}(\varpi,\tilde q)^{-1}}\right\} \ ,
\ee
where $y_0$ is the root of $v(y)$,
\bea
\begin{split}
 I_1(\varpi,\tilde q) &=  \int_0^{y_0} d y \sqrt{-v(y)} = \int_0^{y_0} d y \sqrt{-1 + \varpi^2e^{-2 y} - \tilde q^2 e^{-(2+\delta) y} } \,,\\
 I_2(\varpi,\tilde q) &=\int_{y_0}^\infty d  y \le(\sqrt{v( y)}-1\ri) = \int_{y_0}^\infty d  y \le(\sqrt{1 - \varpi^2e^{-2 y} + \tilde q^2 e^{-(2+\delta) y} }-1\ri)\, ,
 \end{split}
 \eea
 and the reflection amplitude agrees with~\eqref{Refdef} for $\widetilde Q=0$:
 \be
\mathcal{R}(\varpi,\tilde q) =  -\frac{ \Gamma \left(1+i \widetilde S\right) \Gamma \left(\frac{1}{2} \left(-i \varpi -i \widetilde S+1\right)\right)^2}{\Gamma \left(1-i \widetilde S\right)\Gamma \left(\frac{1}{2} \left(-i \varpi +i \widetilde S+1\right)\right)^2 }  \ .
 \ee%
For $\delta = 0$ this reduces to the result found in~\cite{Betzios:2017dol} and in~\eqref{corrfinal}, 
up to the term $i/2$ in the curly brackets which arises due to our different definition of the source term (see Appendix~\ref{app:WKB}). The integrals $I_{1,2}$ can also be evaluated in closed form for $\delta = -1$ ($z=2$), $\delta= -2$ ($z \to \infty$), and $\delta=2$ ($z=1/2$), but the results are mostly lengthy and unilluminating so we do not present them here. The last option appears to violate the null energy condition which requires $z \geq 1$ (when $\theta<0$ which is the case at large $\xi$)~\cite{Giataganas:2017koz}.

The above expression holds in the limit $\xi \to \infty$ in the large-$\varpi$ region of the complex $\varpi$-plane, which includes the region with $\mathrm{Re} \, \varpi -\sqrt{1+\tilde q^2}\gg \xi^{-2/3}$. When $\varpi \to \sqrt{1+\tilde q^2}$ the above expressions fails as $y_0$ becomes small and we need to use the result obtained by matching the near-horizon solution with the Airy functions (see Appendix~\ref{app:smally0matching})~\cite{Casalderrey-Solana:2018uag}. In this case we find that
\bea
\begin{split} \label{GAiry}
 G &= -\frac{1}{2}r_h^{-\xi } \exp\Big[\xi \le( y_0- I_2(\varpi,\tilde q)\right)\Big] \\
 &\times\frac{ \text{Bi}\left(-\frac{y_0 \sqrt[3]{v'(0) \xi ^2}}{2^{2/3}}\right)+\frac{\sqrt[3]{2} i\,  v'(0)^{1/3}}{ \sqrt{2\sqrt{1+\tilde q^2}\delta \varpi}\, \xi ^{1/3}} \frac{\mathcal{R}+1}{\mathcal{R}-1}\,\text{Bi}'\left(-\frac{y_0 \sqrt[3]{v'(0) \xi ^2}}{2^{2/3}}\right)}{ \text{Ai}\left(-\frac{y_0 \sqrt[3]{v'(0) \xi ^2}}{2^{2/3}}\right)+\frac{\sqrt[3]{2} i\,  v'(0)^{1/3}}{ \sqrt{2\sqrt{1+\tilde q^2}\delta \varpi}\, \xi ^{1/3}} \frac{\mathcal{R}+1}{\mathcal{R}-1}\,\text{Ai}'\left(-\frac{y_0 \sqrt[3]{v'(0) \xi ^2}}{2^{2/3}}\right)} \ ,
\end{split}
\eea
where $\delta \varpi = \varpi-\sqrt{1+q^2}$.
This approach, where $\delta \varpi$ is assumed to be small, was used to analyze fluctuations around the geometry dual to Bjorken flow at large D in~\cite{Casalderrey-Solana:2018uag}, which is related to our setup by generalized dimensional reduction. Indeed one can check that their results for the QNMs agree with the poles of~\eqref{GAiry} for $\delta=0$. In Appendix~\ref{app:Bjorken} we check that our results for the QNMs in~\eqref{corrfinal} also apply for the fluctuations around the Bjorken flow (both in the neutral and charged cases), therefore extending the results of~\cite{Casalderrey-Solana:2018uag} to QNMs with lower $\mathrm{Im}\varpi$ and to charged flows.

The region of small $|\varpi|$ in the complex $\varpi$-plane is covered by the solution where $v(y)$ has no roots\footnote{For $\delta>0$ there is also a region where $v(y)$ has two roots. However since this means that $0<z<1$ the null energy condition is violated. We do not consider this case here. } in the integration interval. Therefore the intermediate matching with the Airy functions can be dropped. In this case the correlator becomes
\be
 G =  r_h^{-\xi } e^{-\xi \int_0^{\infty }  \left(\sqrt{v(z)}-1\right) \, dz} \mathcal{R}(\varpi,\tilde q) + \mathrm{sgn}\le(\mathrm{Im}\,\varpi\ri)\frac{i}{2}r_h^{-\xi } \exp\Big[\xi \le( y_0- I_2(\varpi,\tilde q)\right)\Big] \ ,
\ee
which is the counterpart of the expression~\eqref{corrsmallomega} in Sec.~\ref{sec:ancorrs}.
Actually the second discontinuous term (which carries no information on the IR structure) is not obtained from the matching procedure, but was inserted by hand for the result to match with~\eqref{Glargeom}. This reflects the issue due to working with the WKB approximation which we discuss in Appendix~\ref{app:WKB}: the subleading terms of the source solution near the boundary cannot be controlled analytically which leaves the correlator ambiguous (but in a trivial way). The discontinuity of the result at real $\varpi$ is exponentially suppressed with respect to its absolute value.
 

\section{Discussion}

Our analytic results for the QNM spectra {of charged black branes} may have indirect implications for the quark-gluon plasma produced at relatively lower energy heavy ion experiments, such as at RHIC and at the future planned FAIR and NICA, that explore the finite baryon density regime of QCD. In particular, fluctuations of the plasma around the conjectured critical point on the phase diagram could possibly be modelled by the holographic quasi-normal modes we studied here. To turn this into a viable model one should construct an effective theory for the plasma dynamics by coupling the QNMs to the hydrodynamic evolution of the system and run this system with realistic initial conditions e.g. from Pythia, and compute observables such as the behavior of the flow parameters near criticality. The advantage of having analytic control over the modes becomes apparent as one can include as many modes as desired and may consider an attractor solution for the hydrodynamic evolution \cite{Heller:2015dha}. On the other hand the breakdown of the hydrodynamic approximation inflicted by the ``non-decoupled'' modes near criticality and by the ``decoupled'' charged modes near extremality may have an interesting bearing on the observables in strongly coupled plasmas, if not in the QGP. {It would be very interesting to understand the generic features of the approach to thermal equilibrium near criticality.} 

One could ask whether some of these holographic results are universal or they are completely model dependent; a question we leave unanswered in this paper. Another obvious future direction is extending our study into QNMs with different helicity at finite momentum. We do not expect to obtain analytic results even in the critical limit for generic helicity, {except possibly for the ``decoupled" modes in the helicity-zero and one sector discussed in the introduction}, but the salient features such as the breakdown of hydrodynamics may also be visible 
in these spectra. 

Finally, it will be interesting to sharpen the connection between the critical limit of the 5D EMD theory we studied here with the large D limit of charged black branes. In particular, it will be interesting to investigate, in the context of holographic superconductors \cite{Gubser:2008px,Hartnoll:2008vx}, implications of our analytic results for the QNM spectra to the large D limit of such systems, see e.g. \cite{Emparan:2013oza}. 

\acknowledgments

We thank R.~Emparan, B.~Gout\'eraux, C.~Herzog, A.~Jansen, B.~Meiring and J.~Pedraza for useful discussions. The work of U.G. and N.Z. is part of the D-ITP consortium, a program of the Netherlands Organization for Scientific Research (NWO) that is funded by the Dutch Ministry of Education, Culture and Science (OCW). The work of M.J. has been supported 
by an appointment to the JRG Program at the APCTP through the Science and Technology Promotion Fund and Lottery Fund of the Korean Government. It has also been supported by the Korean Local Governments -- Gyeong\-sang\-buk-do Province and Pohang City -- and by the National Research Foundation of Korea (NRF) funded by the Korean government (MSIT) (grant number 2021R1A2C1010834).

\appendix
\section{Matching the UV and IR solutions in the WKB approach} \label{app:WKB}
In this appendix we carry out the matching in the WKB approach considered in Sec.~\ref{sec:WKB}, i.e., we construct the complete IR regular solution to the fluctuation equations in the limit of large $\xi$ by combining the solutions given in~\eqref{WKB},~\eqref{airy}, and~\eqref{horsol}.

We start from the IR. First, regularity of~\eqref{horsol} at the horizon implies
\be \label{reflampl}
 \frac{C_+^\mathrm{IR}}{C_-^\mathrm{IR}} = -\frac{ \Gamma \left(1+i \widetilde S\right) \Gamma \left(\frac{1}{2} \left(-i \varpi -i \widetilde S+1\right)\right)^2}{\Gamma \left(1-i \widetilde S\right)\Gamma \left(\frac{1}{2} \left(-i \varpi +i \widetilde S+1\right)\right)^2 } \equiv \mathcal{R}(\varpi,\tilde q)\, ,
\ee
where we defined the reflection amplitude $\mathcal{R}$.

The next task is to match the IR solution with the WKB approximation. The IR leading corrections to the IR solution are $\sim 1/\xi$ and $\sim |\log w|/\xi$. The corrections to the WKB approximation are $\sim 1/\xi$ and $\sim w$. For large enough $\xi$ there is therefore an overlapping region, $e^{-\# \xi} \ll w \ll 1$, where both formulas work~\cite{Betzios:2017dol}. Taking $w \ll 1$, the IR solution becomes
\be \label{IRUVlimit}
 \phi(w) \simeq C_+^\mathrm{IR}\, w^{\frac{1}{2} \left(1-i \widetilde S-i \varpi\right)}+ C_-^\mathrm{IR}\, w^{\frac{1}{2} \left(1+i \widetilde S-i \varpi\right)} \ .
\ee
Near the horizon (small $y = - \frac{1}{\xi} \log w$) for the relevant parameter values we have $v(y)<0$. Therefore we define 
\bea
 e^{i\omega r} r^{-\xi/2} \phi(y) = \zeta(y) &=& \frac{\widetilde C_+}{|v(y)|^{1/4}} \exp\le(\int_{y_0}^y d\tilde y \frac{\xi}{2} i \sqrt{-v(\tilde y)} \ri) \\
 &&+  \frac{\widetilde C_-}{|v(y)|^{1/4}} \exp\le(-\int_{y_0}^y d\tilde y \frac{\xi}{2} i \sqrt{-v(\tilde y)} \ri) \ .
\eea
At small $y$ (more precisely substituting $y = - \frac{1}{\xi} \log w$ and dropping terms $\sim 1/\xi$) we find that
\bea
 \phi(w) &\simeq&  \frac{\widetilde C_+}{-\tilde q^2+\varpi ^2-1}\, r_h^{\xi/2}e^{-\frac{1}{2}i\xi\varpi- \frac{1}{2} \xi i I_1(\varpi,\tilde q)}w^{\frac{1}{2} \left(1-i \widetilde S-i \varpi \right)} \\
  &&+\frac{\widetilde C_-}{-\tilde q^2+\varpi ^2-1}\, r_h^{\xi/2}e^{-\frac{1}{2}i\xi\varpi + \frac{1}{2} \xi i I_1(\varpi,\tilde q)}w^{\frac{1}{2} \left(1+i \widetilde S-i \varpi \right)} 
\eea
where
\be \label{I1def}
I_1(\varpi,\tilde q) =  \int_0^{y_0} d y \sqrt{-v(y)} = \int_0^{y_0} d y \sqrt{-1 + \varpi^2e^{-2 y} - \tilde q^2 e^{-(2+\delta) y} } \ .
\ee
Comparing to~\eqref{IRUVlimit} we conclude that
\bea \label{firstmatching}
 C_+^\mathrm{IR} &=& \frac{\widetilde C_+}{-\tilde q^2+\varpi ^2-1}\, r_h^{\xi/2}e^{-\frac{1}{2}i\xi\varpi - \frac{1}{2} \xi i I_1(\varpi,\tilde q)} \\
 C_-^\mathrm{IR} &=& \frac{\widetilde C_-}{-\tilde q^2+\varpi ^2-1}\, r_h^{\xi/2}e^{-\frac{1}{2}i\xi\varpi + \frac{1}{2} \xi i I_1(\varpi,\tilde q)}
\eea

Since the WKB approximation fails near $y_0$, we next match it with the solution~\eqref{airy} in terms of the Airy functions. One might expect that the WKB approximation works for $|y-y_0|\gg 1/\xi^2$, but a closer look at the subleading corrections to the approximation reveals that $|y-y_0|\gg 1/\xi^{2/3}$ is required for them to be small. This is however enough for overlap since the Airy solution is fine for $|y-y_0|\ll 1$. Therefore we expand the WKB solution around $y=y_0$:
\be 
 \zeta(y) \simeq \widetilde C_+ \frac{e^{- \frac{1}{3} i \xi  \sqrt{v'(y_0) (y_0-y)^3}}}{v'(y_0)^{1/4} (y_0-y)^{1/4}} + \widetilde C_- \frac{e^{+ \frac{1}{3} i \xi  \sqrt{v'(y_0) (y_0-y)^3}}}{v'(y_0)^{1/4} (y_0-y)^{1/4}} 
\ee
and compare this to the expansion of the Airy solution for $y-y_0 \to - \infty$ (which is valid for $1/\xi^{2/3} \ll |y-y_0|\ll 1$):
\bea
 \zeta(y) &\simeq & \frac{e^{- \frac{1}{3} i \xi  \sqrt{v'(y_0) (y_0-y)^3}}}{(y_0-y)^{1/4}} \left(\frac{\sqrt[3]{-1-i} C_B}{2 \sqrt{\pi } \sqrt[6]{\xi } \sqrt[12]{v'(y_0)} }+\frac{\sqrt[3]{-1+i} C_A}{2 \sqrt{\pi } \sqrt[6]{\xi } \sqrt[12]{v'(y_0)} }\right)\\
 &&+\frac{e^{+ \frac{1}{3} i \xi  \sqrt{v'(y_0) (y_0-y)^3}}}{(y_0-y)^{1/4}} \left(\frac{\sqrt[3]{-1-i} C_A}{2 \sqrt{\pi } \sqrt[6]{\xi } \sqrt[12]{v'(y_0)} }+\frac{\sqrt[3]{-1+i} C_B}{2 \sqrt{\pi } \sqrt[6]{\xi } \sqrt[12]{v'(y_0)} }\right)
\eea
Comparing the expressions we find
\bea \label{secondtmatching}
 \widetilde C_+ &=&\frac{\sqrt[3]{-1+i}\sqrt[6]{v'(y_0)} \, C_A}{2 \sqrt{\pi } \sqrt[6]{\xi } }+\frac{\sqrt[3]{-1-i} \sqrt[6]{v'(y_0)}\, C_B}{2 \sqrt{\pi } \sqrt[6]{\xi } }\\
 \widetilde C_- &=&\frac{\sqrt[3]{-1-i}\sqrt[6]{v'(y_0)} \, C_A}{2 \sqrt{\pi } \sqrt[6]{\xi } }+\frac{\sqrt[3]{-1+i}\sqrt[6]{v'(y_0)} \, C_B}{2 \sqrt{\pi } \sqrt[6]{\xi } }
\eea

The remaining matching task is that between the Airy functions and the WKB approximation near the boundary given in~\eqref{WKB}. It is convenient to first change the lower limits integrations in these definitions from zero to $y_0$, the change can be absorbed in the coefficients $C_\pm$. The matching is analogous to that done above. At small $y-y_0$ the WKB approximation becomes
\be
  \zeta(y) \simeq C_+ \frac{e^{ \frac{1}{3}  \xi  \sqrt{v'(y_0) (y-y_0)^3}}}{v'(y_0)^{1/4} (y-y_0)^{1/4}} + C_- \frac{e^{- \frac{1}{3}\xi  \sqrt{v'(y_0) (y-y_0)^3}}}{v'(y_0)^{1/4} (y-y_0)^{1/4}} 
\ee
which is to be compared to the asymptotics of the Airy functions for $y-y_0 \to \infty$:
\be
 \zeta(y) \simeq \frac{C_A e^{-\frac{1}{3}  \xi  \sqrt{v'(y_0) (y-y_0)^3}}}{2^{5/6} \sqrt{\pi } \sqrt[6]{\xi } \sqrt[12]{v'(y_0)}\sqrt[4]{y-y_0}}+\frac{\sqrt[6]{2} C_B e^{ \frac{1}{3}\xi  \sqrt{v'(y_0) (y-y_0)^3}}}{\sqrt{\pi } \sqrt[6]{\xi } \sqrt[12]{v'(y_0)}\sqrt[4]{y-y_0}} \ .
\ee
We find that 
\be \label{thirdmatching}
 C_- = \frac{C_A\sqrt[6]{v'(y_0)}}{2^{5/6} \sqrt{\pi } \sqrt[6]{\xi }} \ , \qquad C_+ = \frac{2^{1/6}C_B\sqrt[6]{v'(y_0)}}{ \sqrt{\pi } \sqrt[6]{\xi }} \ .
\ee

Finally we need to expand the WKB solutions near the boundary. Switching back to $r$, and recalling the relation $e^{i\omega r} r^{-\xi/2} \phi(r) = \zeta(r)$, we find\footnote{We assume here that $\delta>-2$. The special case $\delta=-2$ can be treated similarly. For $\delta<-2$ an honest UV completion is apparently required.} that
\be \label{UVexpansion}
 \phi(r) \simeq  C_+ r_h^{\xi /2} e^{\frac{\xi}{2}I_2(\varpi,\tilde q)-\frac{\xi  y_0}{2}} + C_-  r_h^{-\frac{\xi }{2}} e^{\frac{\xi  y_0}{2}-\frac{\xi}{2}I_2(\varpi,\tilde q)}r^{\xi }
\ee
where
\be \label{I2def}
I_2(\varpi,\tilde q) = \int_{y_0}^\infty d  y \le(\sqrt{v( y)}-1\ri) = \int_{y_0}^\infty d  y \le(\sqrt{1 - \varpi^2e^{-2 y} + \tilde q^2 e^{-(2+\delta) y} }-1\ri)
\ee

The correlator is then found as the ratio of the coefficients of the vev and source terms. We cannot, however, compute the subleading term $\sim r^\xi$ in the source function because we already took $\xi \to \infty$. Therefore the definition of the correlator remains a bit imprecise. This does not affect the QNMs because redefinitions of the source function by adding or subtracting multiples of the vev term only affects the correlator trivially. Actually, the precise definition of the source function used here is that it exactly (up to an overall coefficient) evolves to the Airy Bi function near $y=y_0$. The result for the correlator is given in Eq.~\eqref{Glargeom}. 

\subsection{Matching at small $y_0$} \label{app:smally0matching}

When $y_0$ is small, one should match the solution near the horizon~\eqref{horsol} directly with the Airy functions~\eqref{airy} rather than using the WKB approximation as an intermediate solution. The direct matching works for $|y_0| \ll 1$, while the matching with the WKB approximation to be valid requires that the term in the argument of the Airy functions is large:  $y_0 v'(0)^{1/3} \xi^{2/3} \gg 1 $. For large enough $\xi$ there is an interval where both matchings work given by $ v'(0)^{-1/3} \xi^{-2/3} \ll y_0 \ll 1$.

We notice that $y_0$ becomes small when $\varpi -\sqrt{1+\tilde q^2} \equiv \delta \varpi \to 0$. In this limit
\be
 y_0 = \frac{\sqrt{1+\tilde q^2}}{1-\tilde q^2 \delta/2} \delta \varpi + \mathcal{O}\le(\delta\varpi^2\ri) \ , \qquad v'(0) = 2-\tilde q^2 \delta + \mathcal{O}\le(\delta\varpi\ri) \ .
\ee

From this discussion we conclude that in particular when $y_0 \sim \xi^{-2/3}$ (or at even smaller $y_0$) the WKB matching fails and the reliable result is obtained through matching with the Airy function. To implement this limit, we define $y_0 = \hat y/\xi^{2/3}$, substitute $y = -\log w/\xi$ and take $\xi \to \infty$ in~\eqref{airy} which yields
\bea \label{phiAiry}
 \phi(w) &=& w^{\frac{1}{2}-\frac{i \varpi }{2}} r_h^{\xi/2} e^{-\frac{1}{2} i \xi \varpi}\Bigg[C_A \text{Ai}\left(-\frac{\sqrt[3]{v'(0)}\hat y}{2^{2/3}}\right)+C_B \text{Bi}\left(-\frac{\sqrt[3]{v'(0)}\hat y}{2^{2/3}}\right)\nonumber\\\nonumber
 &&-\frac{\sqrt[3]{v'(0)}  \left(C_A \text{Ai}'\left(-\frac{\sqrt[3]{v'(0)}\hat y}{2^{2/3}}\right)+C_B \text{Bi}'\left(-\frac{\sqrt[3]{v'(0)}\hat y}{2^{2/3}}\right)\right)}{2^{2/3} \sqrt[3]{\xi }}\log (w)\Bigg]
\eea
whereas near the boundary ($w \ll 1$) and at small $\delta \varpi$ the horizon solution is
\be \label{phihor}
 \phi(w) =(C_-^\mathrm{IR}+C_+^\mathrm{IR}) w^{\frac{1}{2}-\frac{i \varpi }{2}}+ \frac{1}{2} i (C_-^\mathrm{IR}-C_+^\mathrm{IR}) \sqrt{2\sqrt{1+\tilde q^2}\delta \varpi}\, w^{\frac{1}{2}-\frac{i \varpi }{2}} \log (w) \ .
\ee
The relation between the coefficients $C_{A,B}$ and $C_\pm^\mathrm{IR}$ is then found by comparing these two expressions. Further matching with the WKB solution near the boundary (which gives the relations~\eqref{thirdmatching}) we can read off the correlator from the near boundary expansions. The result is given in~\eqref{GAiry}.

\subsection{Matching at small $y_0$: subleading corrections}

It is possible to derive an expression for the coefficients at small $y_0$, i.e., for  $\varpi = \sqrt{1+q^2}$, which improves significantly that obtained from the expressions of Sec.~\ref{app:smally0matching}. This is possible essentially because the corrections for the expression in terms of the Airy functions in~\eqref{phiAiry} are suppressed by powers of $\xi^{-1/3}$ instead of $1/\xi$. Therefore for small $y_0$ we can also determine the subleading correction to~\eqref{airy} by matching with the near horizon solution. This kind of corrections were considered in a slightly different context in~\cite{Casalderrey-Solana:2018uag}.

The relevant subleading corrections to the Airy functions are found by including the second derivative term in the expansion
\be
 v(y) = v'(y_0)(y-y_0)+ \frac{1}{2} v''(y_0) (y-y_0)^2 + \cdots
\ee
and by solving the equation~\eqref{schrode} perturbatively. For the solution giving the UV normalizable mode the result reads
\be
 h(y) = C_A\left[\mathrm{Ai}\le(\Delta\ri) +\frac{ v''(y_0) \Delta (\Delta  \text{Ai}'(\Delta )-\text{Ai}(\Delta ))}{5 \sqrt[3]{2} v'(y_0)^{4/3} \xi ^{2/3}} + \mathcal{O}\left(\frac{1}{\xi^{4/3}}\right)\right] 
\ee
in the limit where
\be
 \Delta = \frac{(y-y_0)\sqrt[3]{v'(y_0)\xi^2}}{2^{2/3}}
\ee
is kept fixed as $\xi \to \infty$ (which will also be the limit relevant for matching). Similar result holds for the nonnormalizable mode: the Airy $\mathrm{Ai}$ function is just replaced by Airy $\mathrm{Bi}$.

What remains to be done is to consider the limit of fixed $w$ at large $\xi$ and match with~\eqref{phihor}. As above we therefore substitute $y = -\log w/\xi$ and develop the Airy function as a series, which gives
\begin{align} \label{Airynlo}
 &w^{-\frac{1}{2}+\frac{i \varpi }{2}}\phi(w) = C_A \Bigg\{\!\le[\mathrm{Ai}\le(\Delta_0\ri) +\frac{ v''(y_0) \Delta_0 (\Delta_0  \text{Ai}'(\Delta_0 )-\text{Ai}(\Delta_0 ))}{5 \sqrt[3]{2} v'(y_0)^{4/3} \xi ^{2/3}} + \mathcal{O}\left(\frac{1}{\xi^{4/3}}\right)\ri]&\nonumber\\
 & \qquad\quad +\Bigg[\!-\frac{\sqrt[3]{v'(y_0)} \text{Ai}'\left(\Delta _0\right)}{2^{2/3} \sqrt[3]{\xi }}+\frac{v''(y_0) \left(\text{Ai}\left(\Delta _0\right)-\Delta _0 \left(\Delta _0^2 \text{Ai}\left(\Delta _0\right)+\text{Ai}'\left(\Delta _0\right)\right)\right)}{10   v'(y_0)\,\xi}&\nonumber\\
 &\qquad\qquad+\mathcal{O}\left(\frac{1}{\xi^{5/3}}\right)
 \Bigg] \log w + \mathcal{O}\le((\log w)^2\ri)\Bigg\} + C_B\Bigg\{ \mathrm{Ai} \rightarrow \mathrm{Bi}\Bigg\}\ , &
\end{align}
where 
\be
 \Delta_0 = -\frac{y_0\sqrt[3]{v'(y_0)\xi^2}}{2^{2/3}} = -\frac{\sqrt[3]{v'(y_0)}}{2^{2/3}}\hat y \ .
\ee
A priori the terms $\propto \log w/\xi$ might seem irrelevant as the functions near the horizon, which we will be matching the Airy functions with, are only known up to corrections $\mathcal{O}(\log w/\xi)$. However the leading constant term in the near horizon solution vanishes in the scaling limit we will consider, an consequently even the terms   $\propto \log w/\xi$ in~\eqref{Airynlo} turn out to be important.

Finally we match this expression with the near horizon solution
\be \label{phihor2}
 w^{-\frac{1}{2}+\frac{i \varpi }{2}}\phi(w) =(C_-^\mathrm{IR}+C_+^\mathrm{IR}) + \frac{1}{2} i (C_-^\mathrm{IR}-C_+^\mathrm{IR}) \sqrt{\varpi^2-q^2-1}\, \log (w)+ \mathcal{O}\le((\log w)^2\ri)
\ee
which determines the higher order corrections to the relations between the coefficients. We could also match the higher powers in $\log w$ but they will not give us any additional information.

Using the improved matching we may write down a more accurate expression than~\eqref{GAiry} for the correlator near $\varpi = \sqrt{1+q^2}$:
\begin{align} \label{GAiryacc}
 G &= -\frac{1}{2}r_h^{-\xi } \exp\Big[\xi \le( y_0- I_2(\varpi,\tilde q)\right)\Big] & \\\nonumber
\times&\Bigg[ \left(
1 +
\frac{v'' \hat y}{10 v' \xi ^{2/3}}\right)\text{Bi}\left(-\frac{\sqrt[3]{v'} \hat y}{2^{2/3}}\right) +  \Bigg(\frac{\sqrt[3]{2} i (\mathcal{R}+1) \sqrt[3]{v'}}{ (\mathcal{R}-1) \widetilde S\sqrt[3]{\xi }}&\\\nonumber
&+\frac{v'' \hat y^2}{10\ 2^{2/3}  v'^{2/3}\xi ^{2/3}}-\frac{i (\mathcal{R}+1) v'' \hat y}{5\ 2^{2/3}   (\mathcal{R}-1) v'^{2/3} \widetilde S\,\xi}\Bigg)\text{Bi}'\left(-\frac{\sqrt[3]{v'} \hat y}{2^{2/3}}\right)\Bigg]&\\\nonumber
\Big/&\Bigg[ \left(
1 +
\frac{v'' \hat y}{10 v' \xi ^{2/3}}\right)\text{Ai}\left(-\frac{\sqrt[3]{v'} \hat y}{2^{2/3}}\right) +  \Bigg(\frac{\sqrt[3]{2} i (\mathcal{R}+1) \sqrt[3]{v'}}{ (\mathcal{R}-1) \widetilde S\sqrt[3]{\xi }}&\\\nonumber
&+\frac{v'' \hat y^2}{10\ 2^{2/3}  v'^{2/3}\xi ^{2/3}}-\frac{i (\mathcal{R}+1) v'' \hat y}{5\ 2^{2/3}   (\mathcal{R}-1) v'^{2/3} \widetilde S\,\xi}\Bigg)\text{Ai}'\left(-\frac{\sqrt[3]{v'} \hat y}{2^{2/3}}\right)\Bigg]
\end{align}
where the derivatives of $v$ are to be evaluated at $y=y_0$. For $\delta = 0$ we have $v'=2$ and $v''=-4$. 
Since this expression works near $\varpi = \pm\sqrt{1+q^2}$, where the line of the CR QNMs starts from, it is expected to give more accurate predictions than~\eqref{Glargeom} for the locations of the CR modes with highest $\mathrm{Im}\varpi$. We have checked this numerically, and found that indeed this is the case for the leading QNM but already for the second QNM the accuracy is roughly the same for both approximations. For the higher modes,~\eqref{GAiryacc} is not useful.

\section{Charged Bjorken flow at large $D$}\label{app:Bjorken}

Notice that not only the charged CR geometry can be seen as a reduction of a D-dimensional charged black brane, but also the Bjorken flow, that was considered in \cite{Gursoy:2015nza}, 
can be related to the large D Bjorken flow  considered in~\cite{Casalderrey-Solana:2018uag} as follows. We fix here $d=4$ and take the $D$ dimensional metric to read
\be\label{Bmetric1}
 \widetilde{ds}^2 = - A(r,\tau) d\tau^2 +2 d\tau dr + S(r,\tau)^2\left(e^{-(\xi-2)B(r,\tau)}d\hat y^2 +e^{B(r,\tau)} \widetilde{dx}_\perp^2\right)
\ee
where $\tau$ is the boost invariant time and $\hat y$ is the rapidity. The extra coordinates of~\eqref{dimregaction} are included in the $\xi-2$ dimensional transverse space. As $dy^2$ is the flat metric, this identification gives the following Ansatz for the dilaton
\be
e^{\phi(r,\tau)} = \left(e^{B(r,\tau )} S(r,\tau )^2\right)^{-\frac{1}{4} \sqrt{\xi -4} \sqrt{\xi -1}}\ .
\ee
The remaining terms in equation~\eqref{dimregaction} then require that the five dimensional metric is given by
\begin{align}
dx^2 &= \left(e^{B(r,\tau )} S(r,\tau )^2\right)^{\frac{\xi -4}{3}}\Big[- A(r,\tau) d\tau^2 +2 d\tau dr & \nonumber\\& \ \qquad + S(r,\tau)^2\left(e^{-(\xi-2)B(r,\tau)}d\hat y^2 +e^{B(r,\tau)} dx_\perp^2\right)\Big]&
\end{align}
where the transverse space is two dimensional. The gauge field equation of motion is solved (both in the five and $D$ dimensional pictures) by
\be
 \partial_r A_\tau(r,\tau) =  F_{r\tau}(r,\tau) =  Q S(r,\tau )^{1-\xi }
\ee
where $Q$ is the charge of the solution.
By substituting this Ansatz  in the Einstein and dilaton equations following from the original action~\eqref{action} we may verify that they are satisfied if the metric components obey the following system of equations:
\begin{align}\label{Bsys1}
 S'' & = -\frac{\xi-2}{4} S \left(B'\right)^2 \, , \\
 S \dot{S}' & = \frac{\xi}{2} S^2- (\xi-2)\dot{S} S' -\frac{Q^2 S(r,\tau )^{4-2 \xi }}{4 (\xi -1)}\, ,  \\
 S \dot{B}' & = -\frac{\xi-1}{2} \left( \dot{S} B'+\dot{B} S'\right) \, ,  \\
 A'' & = - \xi (\xi - 3) - (\xi-2)(\xi-1)\left( \frac{1}{2}\dot{B} B' - 2 \frac{\dot{S} S'}{S^2} \right) +\frac{(3 \xi -5) Q^2 S(r,\tau )^{2-2 \xi }}{2 (\xi -1)} \, ,  \\
 \ddot{S} & =  \frac{1}{2}\dot{S} A'-\frac{\xi-2}{4}\dot{B}^2 S \label{Bsys2} \, , 
\end{align}
where $f' = \partial_r f$, $\dot{f} = \left(\partial_{\tau} + \frac{1}{2} A(r,\tau) \partial_{r} \right) f$, and the mixed derivative terms $\dot f' \equiv \left(\dot f\right)'$. This system agrees for $Q=0$, with eqs. (3.2)-(3.6) of 
~\cite{Casalderrey-Solana:2018uag} that describe the neutral D-dimensional Bjorken flow. 

At the level of quasi-normal modes, we can also observe that, after passing to Eddington-Finkelstein conventions, the fluctuation equation  \eqref{h2eq} for the helicity-two modes coincides with eq. (B.2) of \cite{Casalderrey-Solana:2018uag} 
for the analogous modes of the black brane in the large D limit. 

\subsection{Perturbative analysis in the near-horizon region}
In the limit of large $\xi$ it is possible to find solutions to the system \eqref{Bsys1}-\eqref{Bsys2} as a series in powers of $1/\xi$. First of all we define the new functions
\be
\widetilde{A}\equiv \frac{A}{A_V},\qquad \widetilde{B}\equiv (\xi-2)\left(B-B_V\right),\qquad \widetilde{S}\equiv \frac{S}{S_V},
\ee
where $A_V,B_V,S_V$ describe $AdS_D$ when substituted in the ansatz \eqref{Bmetric1}, and are explicitly given by
\be
A_V=r^2,\qquad B_V=\frac{2}{\xi-1}\log\left(\frac{r}{1+r\tau}\right),\qquad S_V=r^{\frac{\xi-2}{\xi-1}}(1+r\tau)^{\frac{1}{\xi-1}}.
\ee
Next we introduce the new variable
\be
w\equiv r^{-\xi},
\ee
and make the following ansatz for the power series expansion of the tilded functions:
\begin{align}
\widetilde{A}(w,\tau)&=\sum_{n=1}^\infty \frac{1}{\xi^n}A_n(w,\tau)\\
\widetilde{B}(w,\tau)&=\sum_{n=1}^\infty \frac{1}{\xi^n}B_n(w,\tau) \label{Btildeexp}\\
\widetilde{S}(w,\tau)&=\sum_{n=1}^\infty \frac{1}{\xi^n}S_n(w,\tau).
\end{align}
Substituting this ansatz in the equations of motion we find the following equations at lowest order: 
\begin{align}
1-\frac{\mathcal{Q}^2 w^2}{(1+\tau)^2}-A_0+w \partial_w A_0&=0\\
\partial_\tau A_0+\frac{w}{1+\tau}\partial_w A_0 &=0\\
1-A_0+w\partial_w A_0-\frac{w^2}{2}\partial^2_wA_0=0,
\end{align}
where we defined the rescaled charge $Q^2= 2\xi^2\mathcal{Q}^2$.  
The zeroth order solution is then found to be
\begin{align}
A_0(w,\tau)&=1-w\, a^{(1)}_0(\tau)+w^2 a^{(2)}_0(\tau)\\
B_0(w,\tau)&=0\\
S_0(w,\tau)&=1,
\end{align}
with
\be
a^{(1)}_0(\tau)=\frac{1+\mathcal{Q}^2}{1+\tau},\qquad \qquad a^{(1)}_0(\tau)=\frac{\mathcal{Q}^2}{(1+\tau)^2}.
\ee
As expected, $A_0$ therefore is the blackening factor of a shrinking charged black brane.
The solution at NLO can also be found explicitly but it is lengthy and we do not include it here.

\subsection{Perturbative analysis in the near-boundary region}
The charge dependence of the solution is completely absent in the near-boundary region, so the perturbative solution will be exactly the same as the one presented in \cite{Casalderrey-Solana:2018uag}.


\subsection{Non-perturbative modes near the boundary}
Since we are interested in the near-boundary limit, i.e. weak gravity limit, we will consider the following Ansatz\footnote{The coordinate $r$ here will correspond to the inverse of $r$ of the main text.}
\begin{align}
A(r,\tau)&=r^2\left(1-\frac{a(r,\tau)}{r^\xi}\right),\\
B(r,\tau)&=\frac{2}{\xi-1}\log\left(\frac{r}{1+r \tau}\right)+\frac{1}{\xi-2}\frac{b(r,\tau)}{r^\xi},\\
S(r,\tau)&=r\left(\frac{1+r\tau}{r}\right)^{\frac{1}{\xi-1}}\left(1+\frac{s(r,\tau)}{r^{\xi+1}}\right)
\end{align}
and we will   drop terms quadratic or higher in $a,b,s$. Moreover we will write
\begin{align}
b(r,\tau)&=\left(b_0
(r,\tau)+\dots\right)\Omega(r,\tau),\\
a(r,\tau)&=\left(\frac{1}{\xi}a_1
(r,\tau)+\dots\right)\Omega(r,\tau),\\
s(r,\tau)&=\left(s_2
(r,\tau)+\dots\right)\Omega(r,\tau)
\end{align}
and
\be
\Omega(r,\tau)=r^{\xi/2}e^{-i\frac{\xi\varpi}{2}\left(\tau+\frac{1}{r}\right)}F(r),
\ee
where $F(r)$ is a function to be determined. 
In~\cite{Casalderrey-Solana:2018uag} the solution was found at any $\tau$ and at large $\xi$ by employing a WKB approach.
However for our purposes it is convenient 
to find solutions at late times $\tau\rightarrow\infty$ without expanding in powers of $1/\xi$. In this case we may set $F(r)=1$ without loss of generality. A solution is given by
\begin{align}
b_0(r)&=c_+ J_{\frac{\xi}{2}}\left(\frac{\xi \varpi}{2r}\right)+c_- J_{-\frac{\xi}{2}}\left(\frac{\xi \varpi}{2r}\right)\\
a_1(r)
&=s_2(r)
=0
\end{align}
with $c_+$, $c_-$ 
integration constants. Notice that this solution is consistent with the result \eqref{bdrysol} when $k=0$.

\subsection{Non-perturbative modes near the horizon}

The nonperturbative modes near the horizon may be solved analogously by adding a term (small perturbation)
\be
 \Delta \widetilde B(w,\tau) = e^{-i\frac{\xi\varpi}{2}(1+\tau)}Z(w/(1+\tau))
\ee
in~\eqref{Btildeexp} (perturbations of the other components $\widetilde A$ and $\widetilde S$ may be set to zero). Inserting this to the Einstein equations and linearizing gives the condition
\be
(1-\tilde w) \left(1-\mathcal{Q}^2 \tilde w\right) Z''(\tilde w)+\left(\mathcal{Q}^2 (2 \tilde w-1)+\frac{ i \varpi }{\tilde w}-1\right) Z'(\tilde w)-\frac{i \varpi  Z(\tilde w)}{2 \tilde w^2} =0 
\ee
with $\tilde w = w/(1+\tau)$.
This equation agrees with~\eqref{NHflucteq} at $q=0$ and after changing to the Eddington-Finkelstein coordinates. The solutions are given in~\eqref{1F2sols} and in~\eqref{1F2sols2}. 

We have therefore shown that the nonperturbative fluctuation equation match with the fluctuations of the static system both near the boundary and near the horizon. It is immediate that the spectrum of the nonperturbative mode also matches with the QNMs, and in particular it is determined at leading order in $1/\xi$ as the poles of~\eqref{corrsmallomega} and~\eqref{corrfinal}. This extends the analogous result in~\cite{Casalderrey-Solana:2018uag}, which was found in the vicinity of $\varpi=1$, to all frequencies. 

\section{WKB analysis of the transverse fluctuations} \label{app:WKBhighq}

We study the fluctuation of the transverse spin-2 modes in the WKB approximation, extending to the charged case the analysis performed in  \cite{Fuini:2016qsc} for $N=4$ SYM and in \cite{Betzios:2018kwn} for the uncharged plasma. Notice that the WKB approximation in this Appendix means taking the limit of large momentum, and is therefore distinct from the analysis of Sec.~\ref{sec:WKB}.

The equation for the transverse spin 2 fluctuations is given in \eqref{h2eq}, which we can rewrite in the Eddington-Finkelstein coordinates as  
\be \label{h2eqEF}
f(r) H''(r) + (B(r) + 2 i \omega) H'(r) - \left(k^2 + i \omega \frac{f'(r)-B(r)}{f(r)} \right) H(r) = 0 \,, 
\ee
where $$B(r) = \frac{f(r) - \xi + \xi \tilde Q^2 \left( \frac{r}{r_h} \right)^{2(\xi-1)}}{r} \,,$$ 
and $f(r)$ given by \eqref{metricCR}. 
The position of the horizon $\hat r_h$ can be factored out of the fluctuation equations by rescaling the frequency and momentum: 
\be \label{rescaledpom}
 q = \frac{k}{2 \pi T} = \frac{2 k\ell' \hat r_h}{\xi}\ , \qquad \varpi = \frac{\omega}{2 \pi T} = \frac{2 \omega\ell' \hat r_h}{\xi} \ .
\ee
In order to put the equation in the Schr\"odinger form, we redefine the coordinate as  
\be \label{wdef}
 w =\left(\frac{\hat r}{\hat r_h}\right)^\xi \ ,
\ee
and the fluctuation as $H(r) = e^{g(w)} h(w)$. 
It turns out that $g$ has to satisfy 
\be
2 f w'(r)^2 g' + f w''(r) + (B + 2 i \omega )w'(r) = 0 \, \nonumber
\ee 
i.e., 
\be
g' = \frac{(1+ \frac{\xi}{\xi-2} \tilde Q^2) w - (1 + \frac{\xi}{\xi-2}) \tilde Q^2 w^{2-{2\over \xi}} - i \varpi w^{1\over \xi}}{2 w f} \,.
\ee
The equation~\eqref{h2eqEF} takes the form  
\bea\label{eqSchr}
 -h''(w) &+ &V(w) h(w) =  0 \\
 V(w) & =&  g'^2 - g'' + \frac{k^2 + i \omega \frac{f'(r)-B}{f}}{f w'(r)^2}  =  \nonumber \\
&=& -\frac{1}{4 w^2 f^2} \left( w^2 \left(1 + {\xi \tilde Q^2\over \xi-2} (2-q^2) +  {\xi^2  \tilde Q^4 \over (\xi-2)^2}\right) + \right. \nonumber \\
&& \left.   w^{2 \over \xi} (\varpi^2-q^2) + w^{1+{2 \over \xi}} q^2 (1+ {\xi \tilde Q^2 \over \xi -2}) \right.  \\ \nonumber
&& \left. -4  {(\xi-1) \tilde Q^2 \over \xi} w^{2-{2\over \xi}} -8 w^{3-{2\over \xi}} {\tilde Q^2 (\xi-1) \over \xi(\xi-2)} \left(1+ {\tilde Q^2
\xi \over \xi-2}\right) + 4 w^{4-{4\over \xi}} {\tilde Q^4 (\xi-1) \over (\xi-2)^2} \right)
\eea
{This Schr\"odinger form differs from~\eqref{eq:Sform} as we are using a different coordinate and we did not require a specific normalization for the dependence on frequency.}
Near the horizon we find that
\be
 V(w) = -\frac{(1-\tilde Q^2)^2 +\varpi^2}{4(1-\tilde Q^2)^2 (1-w)^2} + \mathcal{O}\left({\frac{1}{1-w}}\right)
\ee
which (combined with the factor $e^g$) gives the behavior $H \sim \mathrm{const.}$ for the ingoing mode and $H \sim (1-w)^{i\varpi/(1-\tilde Q^2)}$ for the outgoing mode.


The boundary conditions for $h(r)$ are the following: 
\begin{eqnarray} 
h(w \sim 0) & \sim& C_1 \left(1+ {\cal O}(w^{1/\xi})\right) + C_2 w \left(1+ {\cal O}(w^{1/\xi})\right) \,, \nonumber\\
h(w \sim 1) &\sim& (1-w)^{{1\over 2}(1 - i {\varpi \over 1- \tilde Q^2})} \,.
\end{eqnarray}  
We introduce $s = \varpi/q$, and we anticipate that in the large $q$ limit the leading behavior is $s = 1 + s_\alpha q^{- \alpha}$, for some value of $\alpha$. 
We can write the potential as 
\bea
V & = & q^2 Q_0 + q^{2-\alpha} Q_\alpha + Q_2 + \ldots  \\
Q_0 & = &  - {1 \over 4 f^2} \left( w^{{2 \over \xi}-1} (1+ {\tilde Q^2 \xi \over \xi-2}) - {\tilde Q^2 \xi \over \xi-2} \right) \,,  \nonumber\\
Q_\alpha & = &  - {s_\alpha \over 2}  {w^{{2\over \xi}-2} \over  f^2} \,,  \nonumber\\
Q_2 &= & {\tilde Q^2 (\xi-1) \over \xi} {w^{-{2\over \xi}} \over f^2} + \ldots  \nonumber
\eea
We see that the term $Q_\alpha$ is subleading in $q$ but it dominates close to the boundary. The term $Q_2$ becomes singular at the boundary as well, but is subdominant compared to $Q_\alpha$. 
%
These expressions show that the charge modifies the coefficients but not the power of the leading singularities. 
For the WKB approximation we take an Ansatz of the form 
\be
h = A \, e^{(q T_0 + q^{1-\alpha} T_\alpha + \ldots)} \,.
\ee
Plugging in the equation and expanding in $q$ we find the conditions  
\be
 A = Q_0^{-1/4} \,, \quad T_0' = \pm Q_0^{1/2} \,, \quad T_\alpha' = {Q_\alpha \over 2 Q_0^{1/2}} \,.
 \ee
The boundary condition at the horizon is imposed by choosing the appropriate sign in the solution for $T_0$.

For small $w$, the WKB expansion reduces to 
\be 
h \sim w^{\xi-2 \over 4 \xi} \, \textrm{exp} \left( {i \xi \over \xi+2} \gamma \, q  w^{\xi+2 \over 2 \xi} + {i s_\alpha \over \gamma (\xi-2)}  q^{1-\alpha} w^{2-\xi \over 2 \xi} \right) \,,
\ee
where $\gamma =\sqrt{1+ {\tilde Q^2 \xi \over \xi-2}}$.
The change of variable $y = w q^{2 \xi \over \xi+2}$ applied to the near-boundary equation and the WKB solution shows that the solution is consistent if 
\be 
\alpha = {2 \xi \over \xi +2 } \,. 
\ee
We obtain then the same asymptotic damping as in the uncharged case.  The coefficient $s_\alpha$ however is modified. 

\bibliographystyle{JHEP}
\bibliography{biblioCQNM}

\end{document}